\documentclass[bookmarks=true,pdfnewwindow=true,colorlinks=true,linkcolor=xlinkcolor,citecolor=xlinkcolor,filecolor=xlinkcolor,urlcolor=xlinkcolor,final=true,,colorlinks=true,inkcolor=purple,citecolor=navyblue,filecolor=cyan,urlcolor=magenta]{aastex631}
\usepackage{cancel}

\usepackage{parskip}
\usepackage{graphicx}
\usepackage{nicefrac}
\usepackage{amsmath}
\usepackage{booktabs}
\usepackage{mwe} 
\usepackage{diagbox}
\usepackage{float}
\usepackage{gensymb}
\usepackage{txfonts}
\usepackage{subfigure}
\usepackage{multirow}
\usepackage[percent]{overpic}
\usepackage{xcolor}

\let\oldput\put
\def\put(#1,#2)#3{%
  \oldput(#1,#2){\sffamily #3}%
}
\definecolor{navyblue}{RGB}{0,50,250}
\definecolor{purple}{RGB}{100,0,250}

\newcommand{\magarc}{mag arcsec\ensuremath{^{\mathrm{-2}}}}
\newcommand{\mulim}{\ensuremath{\mu_{\rm lim}}}
\def\Msun{\ifmmode{\mathrm M_\odot}\else{M$_\odot$}\fi}

\newcommand{\mubreak}{\ensuremath{\mu_\mathrm{brk III}}}
\newcommand{\hr}{\ensuremath{h_{\mathrm{r}}}}
\newcommand{\hz}{\ensuremath{h_{\mathrm{z}}}}

\newcommand{\mui}{\ensuremath{\mu_{0,\mathrm{i}}}}
\newcommand{\mug}{\ensuremath{\mu_{\mathrm{g}}}}
\newcommand{\mur}{\ensuremath{\mu_{\mathrm{r}}}}

\newcommand{\muzero}{\ensuremath{\mu_{0}}}

\def\ugc{UGC\,11859}
\def\df1{UGC\,11859-DF1}

\newcommand{\HI}{\ion{H}{1}}
\shorttitle{Flares, warps, truncations, and satellites in the ultra-thin galaxy \ugc}
\shortauthors{Ossa-Fuentes et al.}
\graphicspath{{./}{figures/}}

\begin{document}

   \title{Flare, Warp, Satellite, and Truncation: the ultra-thin galaxy \ugc}

\correspondingauthor{Alejandro S. Borlaff}
\email{a.s.borlaff@nasa.gov, asborlaff@gmail.com}

\correspondingauthor{Luis Ossa-Fuentes}
\email{luis.ossaf@postgrado.uv.cl}

\author{Luis Ossa-Fuentes}
\affiliation{Instituto de F\'{i}sica y Astronom\'{i}a, Facultad de Ciencias, Universidad de Valpara\'{i}so, Gran Breta\~{n}a 1111, Playa Ancha, Valparaíso, Chile}
\affiliation{Valencian International University (VIU) Valencia, Spain}

\author{Alejandro S. Borlaff}
\affiliation{NASA Ames Research Center, Moffett Field, CA 94035, USA}
\affiliation{Bay Area Environmental Research Institute, Moffett Field, California 94035, USA}
\affiliation{Kavli Institute for Particle Astrophysics \& Cosmology (KIPAC), Stanford University, Stanford, CA 94305, USA}

\author{John E. Beckman}
\affiliation{Instituto de Astrof\'{i}sica de Canarias, C/ V\'{i}a L\'actea, E-38200 La Laguna, Tenerife, Spain}
\affiliation{Facultad de F\'{i}sica, Universidad de La Laguna, Avda. Astrof\'{i}sico Fco. S\'{a}nchez s/n, 38200, La Laguna, Tenerife, Spain}

\author{Pamela M. Marcum}
\affiliation{NASA Ames Research Center, Moffett Field, CA 94035, USA}

\author{Michael N. Fanelli}
\affiliation{NASA Ames Research Center, Moffett Field, CA 94035, USA}



\begin{abstract}

 {The structure of the outskirts of galaxies provides valuable information about their past and evolution. Due to their projected orientation, edge-on isolated galaxies effectively serve as test labs in which to study the three-dimensional structures of galaxies including warps and flares, and to explore the possible sources of such distortions.}
   {We analyzed the structure of the apparently isolated edge-on ultra-thin galaxy \ugc\ to look for the presence of distortions.}
   {The deep optical imaging observations ($\mulim=30.6$ and 30.0 \magarc\ in the $g$ and $r$-bands, resp.) we acquired with the 10.4\,m Gran Telescopio Canarias (GTC) are used to derive the radial and vertical surface brightness profiles and \emph{g-r} color radial profile.}
   {We find that \ugc's disk displays a significant gravitational distortion. A warp is clearly detected on one side of the disk, and the galactic plane on both sides of the centre shows increasing scale height with increasing galactocentric radius, indicating the presence of a flare in the stellar distribution. The surface brightness profile of the disk shows a sharp break at 24 kpc galactocentric radius, and a steep decline to larger radii, an "edge-on truncation", which we associate with the presence of the flare. The present study is the first observational support for a connection between truncations and flares. Just beyond the warped side of the disk a faint galaxy is observed within a small angular distance, identified as a potential interacting companion. Based on ultra-deep g and r photometry we estimate that if the potential companion is at the same distance as \ugc, the stellar mass of the satellite galaxy is $\log_{10}$($M_\odot$) $=6.33^{+0.02}_{-0.02}$.}


\end{abstract}


\section{Introduction}

The cumulative processes of galaxy formation and evolution are imprinted within the dim light that defines the outskirts of galactic stellar disks. At high galactocentric distances, the remains of past and present gravitational interactions remain for several Gyrs, often revealing asymmetries not detectable at the brighter inner regions \citep{MartinezDelgado2009ApJ692955M, Martin2013ApJ77680M}. For this reason, deep imaging explorations of galaxies are a unique tool to reveal the mechanisms that created the Hubble Sequence as observed in the Local Universe. Radial surface brightness profiles are valuable tools for examining the overall structure of disk galaxies \citep{Erwin2008}. The outer zones of the profiles give the most information about the disk structure: azimuthal variations can trace arm structure while the mean radial profile characterizes the outer disk where the surface brightness is reduced. Surface brightness profiles of disk galaxies generally display an exponential decline of intensity with galactocentric radius \citep["Type I",][]{Freeman1970}, while for a non-negligible fraction of disks, the decline steepens at a well defined radius, the "break radius", giving a double exponential profile \citep["Type II", or broken profiles]{Erwin2005}. Early observations were focused on edge-on galaxies, because the foreshortening of the disk aspect gives rise to higher observed brightness, with higher signal-to-noise ratios (SNR) at large galactocentric radii. But the foreshortening also enhances the optical depth of the dust absorption in the disk plane, and compresses the effects of small scale features, making the analysis more difficult \citep{Mathis1990ARA&A..28...37M}.

Type-II profiles may provide critical information to discern different disk formation scenarios. For example, abrupt truncations might reflect the maximum angular momentum  of the initially collapsing cloud \citep{10.1093/mnras/126.6.553}, or, suggest the presence of a dynamically-driven critical star formation threshold for a thin gas disk of galaxy dimensions \citet{Toomre1964, Kennicutt1989}. The transition between galactic and extragalactic magnetic field morphology and strength has also been invoked \citep{Battaner2002} to account for the truncations. Sharp truncations on edge-on disks tend to appear at radii of $\sim$4 inner disk scale-lengths ($h$) from the center of the galaxy \citep{vanderkruit1979, vanderKruit1981a, vanderKruit1981b,Kregel2002MNRAS.334..646K}. In a more recent article \citet{MartinNavarro2012} combined optical images from the Sloan Digital Sky Survey (SDSS) and near infrared images from the Spitzer Survey of Stellar Structure in Galaxies (S4G) to produce radial profiles of 34 near-edge-on galaxies. Their analysis revealed Type~II breaks at an average radius of $\sim$8 kpc and a second, sharper truncation at $\sim$14 kpc, concluding that two different physical mechanisms were responsible for the truncations and the breaks. They suggested that the breaks were associated with thresholds in the star formation efficiency and the sharper truncations detected at earlier epochs by \citet{vanderkruit1979, vanderKruit1981a, vanderKruit1981b} correspond to the foreshortened breaks \citep{MartinNavarro2012}. 

Flares \citep[an increase of the vertical scale length of the disk with increasing galactocentric radius, ][]{Sancisi1979, OBrien2010, Zschaechner2012}, warps \citep[misalignments between inner and outer parts of the disk, ][]{Brinks1984,Bottema1987,Kuijken2001} and other asymmetries can distort the light distribution at the very low surface brightness limits, invisible at the pixel-to-pixel image level, but creating  systematic observational effects in the integrated surface brightness profiles. \citet{Pohlen2002} found that in face-on galaxies, no sharp truncations are observed, and that the profiles were all either Type~I, or Type~II. Further works \citep{Pohlen2004} suggested that flares may create apparent truncated profiles in edge-on galaxies that would not be visible if viewed face-on, explaining previous observations. \citet{Borlaff2016} confirmed that flares can create the effect of surface brightness breaks in edge-on galaxies. These breaks would not be visible in face-on galaxies, explaining the dichotomy between the observations of face-on and edge-on galaxies, and reproducing the observed weakening of the breaks at higher distances from the galactic plane \citep{Pohlen2007}. 

On the other hand, if the apparent truncations found in edge-on galaxies are the foreshortened observations of actual Type~II profiles, we need to find scenarios which can produce the latter. \citet{Pohlen2006} found that profiles in the two photometric bands, $g$ and $r$, are progressively bluer out to the break radius, and then become radially redder. This behaviour was used by Roskar in a series of articles beginning with \citet{Roskar2008} to explain the formation of Type~II disks by a combination of continued star formation due to the sustained infall of low metallicity gas to the disk and the migration of stars outwards caused by the dynamical inhomogeneities of bars and spiral arms. These models predict that the radius of the break between inner zone of lower slope and the outer zone of steeper slope  migrates radially outwards with time. This prediction was confirmed by \citet{Azzollini2008}, who used HST images in the GOODS-South field \citep{Giavalisco2004} to show that break radii have increased systematically by a factor of 1.3 since redshift $z = 1$. The agreement between \citet{Roskar2008} inside-out disk models and the observations suggest that interpreting the observed truncations as foreshortened breaks is valid, negating the need for models aimed at producing sharp truncations found in edge-on galaxies. To clarify the difference between the physical truncations found by \citet{MartinNavarro2012} and the observed truncations in the surface brightness of edge-on galaxies of the type we have observed in \ugc\ we will refer to the latter as an “edge-on truncation”.

\ugc\ is a relatively isolated galaxy\footnote{NASA/IPAC Extragalactic Database: \url{http://ned.ipac.caltech.edu}} ($\alpha=329.530^\circ$, $\delta=+1.0089^\circ$, $D = 51.40$ Mpc following \citet{2010Ap&SS.325..163P}) with a very thin disk oriented edge-on with respect to us. An interesting fact about this galaxy is that despite being relatively isolated, it presents a relatively high asymmetry index in the HI profile \citep{1983AJ.....88..272H,2011yCat..35320117E}. This could indicate possible invisible interactions with the medium at low surface brightness, either with low-brightness satellites, or with intergalactic gas flows, which would be completely invisible to current data. 

In very low surface brightness regions at the limits of the disk,   remnants of past evolutionary processes such as tidal disruptions or the infall of satellite dwarf galaxies can be found. Recent observations of dwarf galaxies in outer, less dense and isolated regions of clusters \citep{Venhola2017A&A608A142V} suggest that these systems had an evolutionary origin related to environments with high angular momentum, feedback induced gas outflows and tidal interactions. Therefore it is an interesting prospect to find out using deep observations of \ugc\ if satellites may be present in similar environments where warps and flares appear and if these features could be related or not to the structure and evolution of the main galaxy.

Obtaining a very deep image of an edge-on “isolated” galaxy with especially low sky background required the use of the 10.4\,m GTC, currently the largest optical/infrared telescope. The data provided opportunities to examine not only the surface brightness profile at the edge of the disk, but also to see whether the galaxy is truly isolated, and whether its disk is truly axisymmetric. During the course of the analysis, we found a faint and previously-undetected possible companion. Additionally, a warp was detected in the stellar disc population distribution, also at very low surface brightness, and probably connected with the companion. We also found clear evidence that the disk is flared. The results of our study therefore resulted in a bounty of findings that we had not anticipated. The methodology of this work is detailed in Sec.\,\ref{Sec:Methods}. The results are presented in Sec.\,\ref{Sec:Results}. We discuss the results in Sec.\,\ref{Sec:Discussion}. The final conclusions can be found in Sec.\,\ref{Sec:Conclusions}. We assume a concordance cosmology \citep[$\Omega_{\mathrm{M}} = 0.3,\Omega_{\mathrm{\Lambda}}=0.7, H_{0}=70 $ km s$^{-1}$ Mpc$^{-1}$, see][]{WMAP2007}. All magnitudes are in the AB system \citep{Oke1971} unless otherwise noted.

\vspace{1cm} 
\section{Methods}
\label{Sec:Methods}

\begin{figure*}[]
 \begin{center}
\centering
\includegraphics[trim={5 0 10 0}, clip, width=\textwidth]{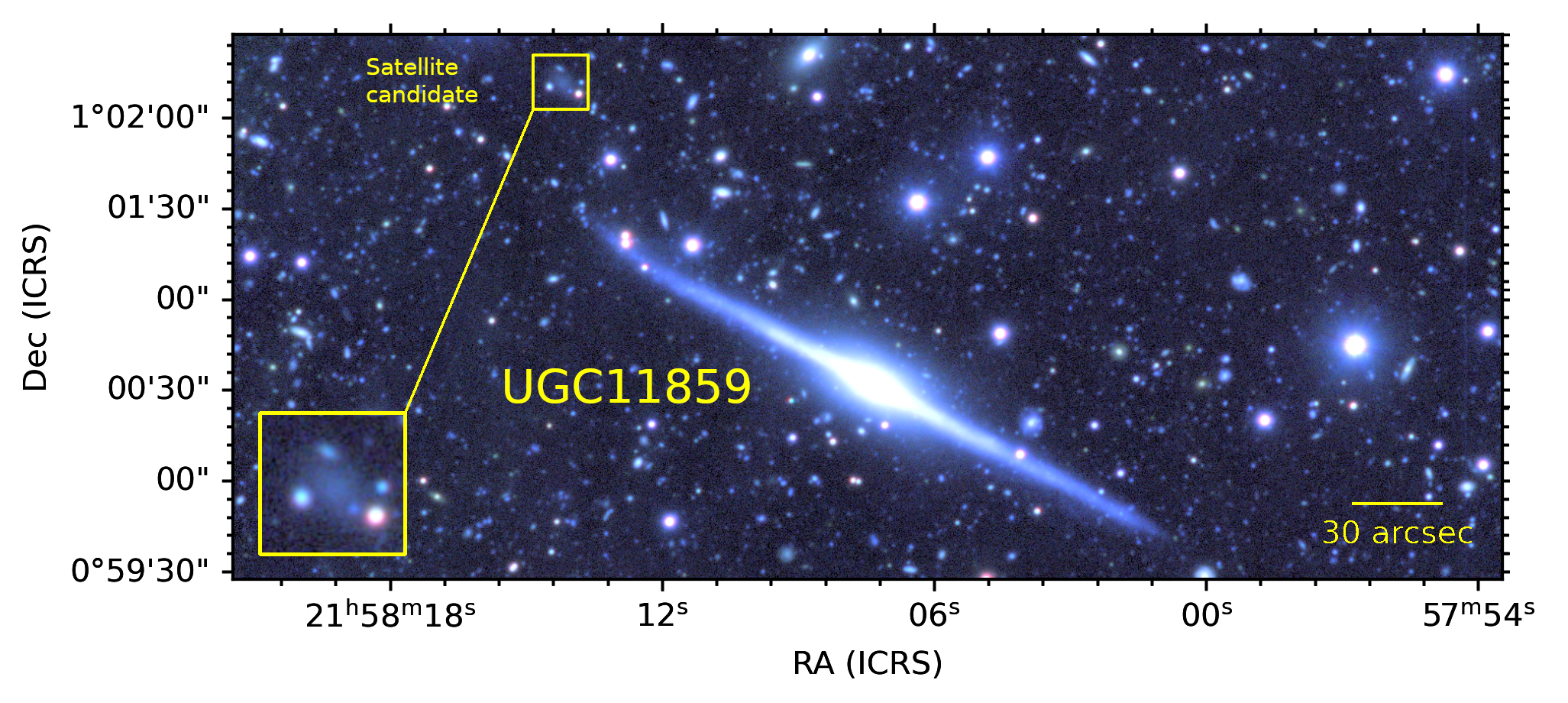}
\caption{Luminance-RGB image of \ugc, generated using the $g$ and $r$ mosaics from GTC (blue, green), and the Stripe 82 deep mosaics (red). The yellow square shows a zoom of the potential satellite galaxy at the northeast of \ugc.}
\label{fig:UGC11895_satelite}
\end{center}
\end{figure*}


\subsection{Observations}
\label{Subsec:methods_observations}
Optical imaging observations of the galaxy \ugc\ were obtained during two hours of observation time in October 2018 with the Gran Telescopio Canarias (GTC) at the Roque de los Muchachos observatory, in La Palma. The observations were made using the OSIRIS camera with SDSS $g$ and $r$ wavelength filters in order to maximize the transmission of the atmosphere and filter system. The field of view was sufficient to simultaneously map the stellar disk and its potential low surface brightness external structures. The $g$ and $r$ photometry provide mass density distribution measures of the object, following the method described in \citet{Bakos2008}. A total of 82 independent exposures were obtained (see Fig.\,\ref{fig:UGC11895_satelite}, the composite image of \ugc). Individual exposures were initially processed using standard calibration products (bias correction, dark, and twilight flats), followed by calibration steps to ensure the suppression of systematic biases related to the uniformity of the background. We detail these steps in the following sections.

\subsection{Image reduction}
\label{Subsec:methods_reduction}

The science and twilight flat exposures were first corrected from bias effects, using a median bias frame per night. At the same time, masks of saturated pixels were generated and then used for the foreground stars stray-light correction described in Sec.\,\ref{Subsec:methods_straylight}. The calibration of the spatial variation of the CCD sensitivity utilized an iterative sky flat-fielding method. The objective of this method is to remove any systematic gradients present in the data products that might contaminate the extended signal on our final mosaics.  

First, we generate a median twilight flat per night and filter, which is used to correct the bias pre-calibrated frames as a first order correction to the flat-field. After this step, significant residuals remain on the images which must be removed by using sky-flat fields. To generate the sky-flats, all individual exposures are normalized to a median value of one and then combined using a robust bootstrapping median algorithm. The result is a set of four second order flat field corrections (one per night and filter). 

The previous process was then further improved by masking astronomical sources in the corrected images, reducing the gradients in the calibration files. We mask all the frames using {\textsf{Gnuastro}}'s {\tt{Noisechisel}} \citep{gnuastro, noisechisel_segment_2019}, employing the same configuration as in \citet{Borlaff2019}. Using the source masks, we repeat previous steps to obtain a second approximation to the sky flat. This process can be repeated until the delta-sky flat corrections converge and the combined frames show no significant structure apart from statistical noise. For this dataset, we achieve convergence in three sky-flat iterations. The photometric calibration of mosaic images was carried out using segmentation catalogs generated with Source Extractor \citep{Bertin1996} and custom software written in \textsf{Python}. We calculate a linear model between the SDSS photometry \citep{York2000} and those obtained with the GTC for independent apertures defined by the Source Extractor segmentation maps. The images were subsequently aligned using the astronomical combination software SWarp \citep{Bertin2010}.

\subsection{Astrometric and photometric calibration}
\label{Subsec:methods_calibrations}

Astrometric calibration and distortion corrections of the individual exposures were performed with \textsf{Scamp} \citep{Bertin2006}. We used the generated source catalogs per exposure, with the SDSS~Data~Release~9 as reference \citep{Eisenstein2011}. \textsf{Scamp} uses the reference and target image catalogs to generate a new WCS header with the corresponding projection distortion parameter. We replace this header in each one of the fits files, and reproject the frames to a common grid with \textsf{Swarp} \citep{Bertin2002}.

The $g$ and $r$ individual images were then calibrated using the photometry data from the IAC~Stripe\,82 photometry \citep{IACStripe82,Roman2018}. This survey obtained images sensitive to low luminosity detection, reaching two magnitudes deeper than the SDSS survey \citep{Trujillo2016}. The project tracks a 2.5\,degree band along the celestial equator (coordinates: $-50^\circ < \alpha < 60^\circ$, $-1.25^\circ < \delta < 1.25^\circ$) reaching a total area of 275 square degrees using five SDSS filters ($ugriz$). In order to calibrate the images, we used {\textsf{Gnuastro}}'s {\tt{Noisechisel}} to create a catalog of detected sources on the Stripe\,82 calibrated mosaics, and measured the flux in the same apertures on the GTC mosaics. We fit a linear calibration correction, avoiding saturated objects (45\,000 electrons on a full well capacity of 65\,535 for the Marconi CCD44-82 detectors of GTC/OSIRIS), and transformed the units on the GTC mosaics. The captured signal (counts) for each pixel in the image was transformed into surface brightness units (\magarc) using:

\begin{equation} \label{eq:Stripe_82}
 \mu =-2.5\cdot\log_{10}(F) + 25.3679
\end{equation}


where $\mu$ is the surface brightness and $F$ the flux per pixel. For OSIRIS/GTC, the pixel scale is 0.256 arcsec px$^{- 1}$. We measure a surface brightness limit in the center of the FOV of $\mulim=30.6$ and 30.0 \magarc\ in the $g$ and $r$ final mosaics, measured as a 3$\sigma$ detection over $10\times10$ arcsec$^2$.

\subsection{Foreground stray-light correction}
\label{Subsec:methods_straylight}

\begin{figure*}[]
 \begin{center}
\centering
\includegraphics[trim={0 35 65 0}, clip, width=0.4835\textwidth]{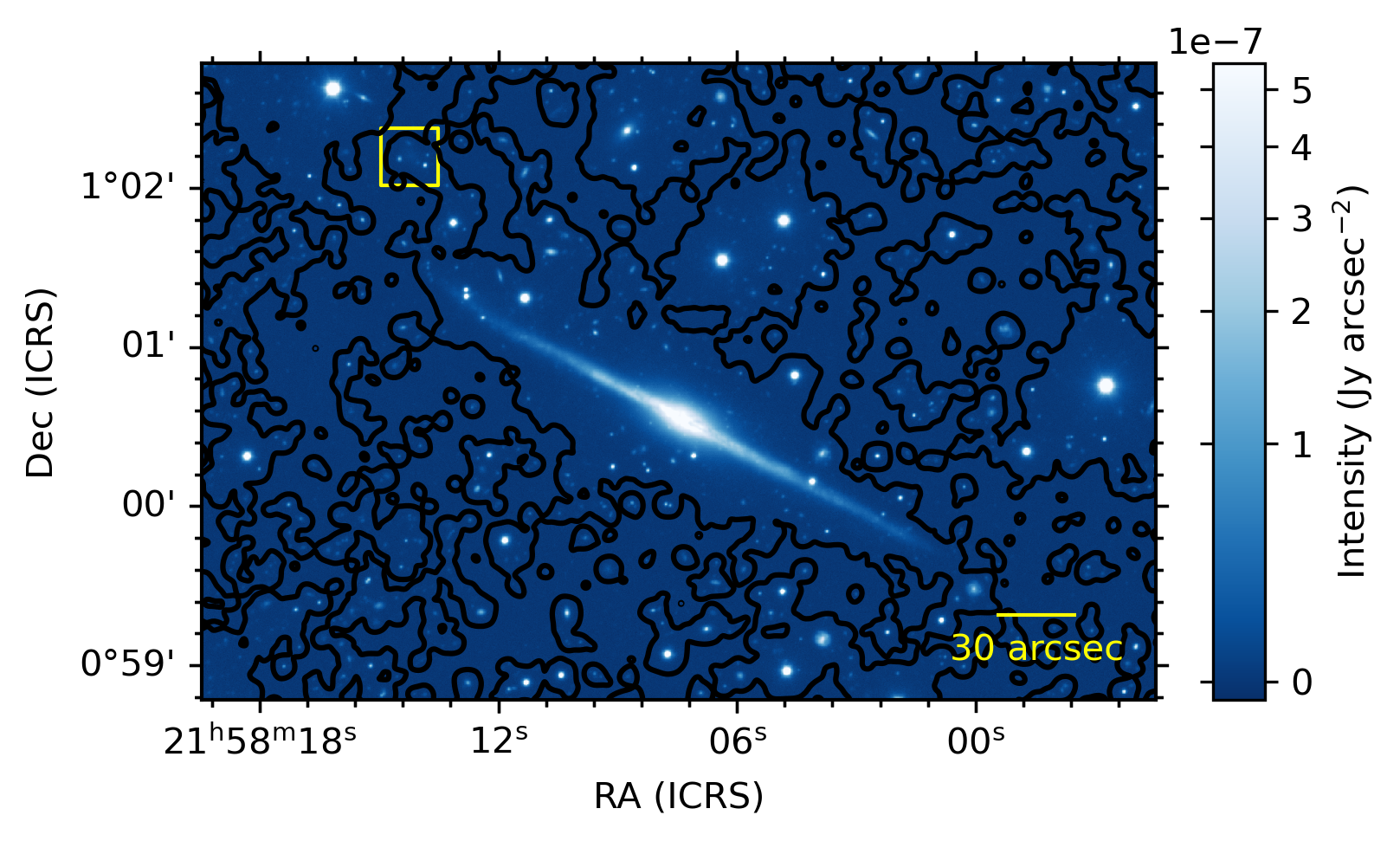}
\includegraphics[trim={50 35 0 0}, clip, width=0.5065\textwidth]{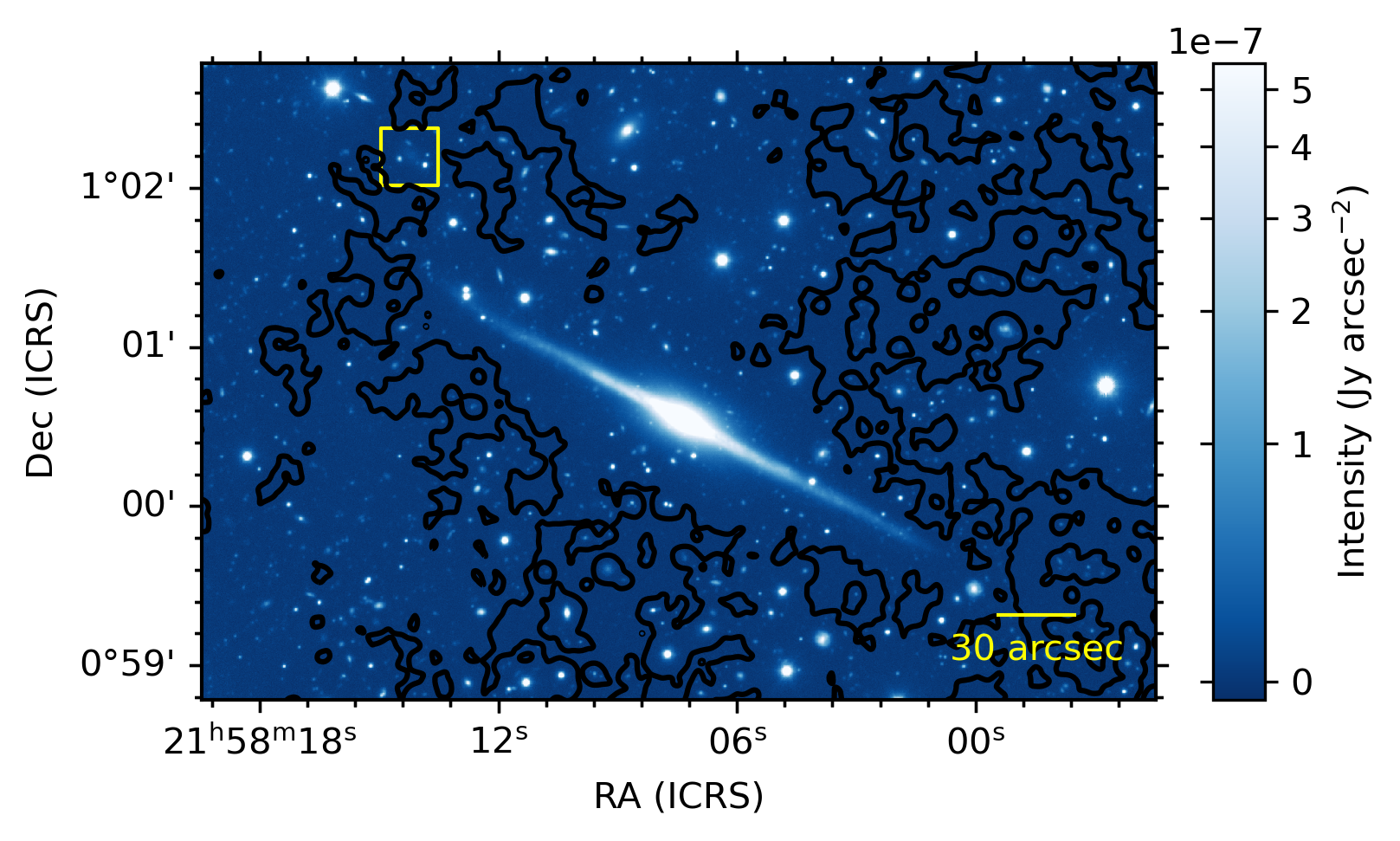}

\includegraphics[trim={0 35 65 0}, clip, width=0.4835\textwidth]{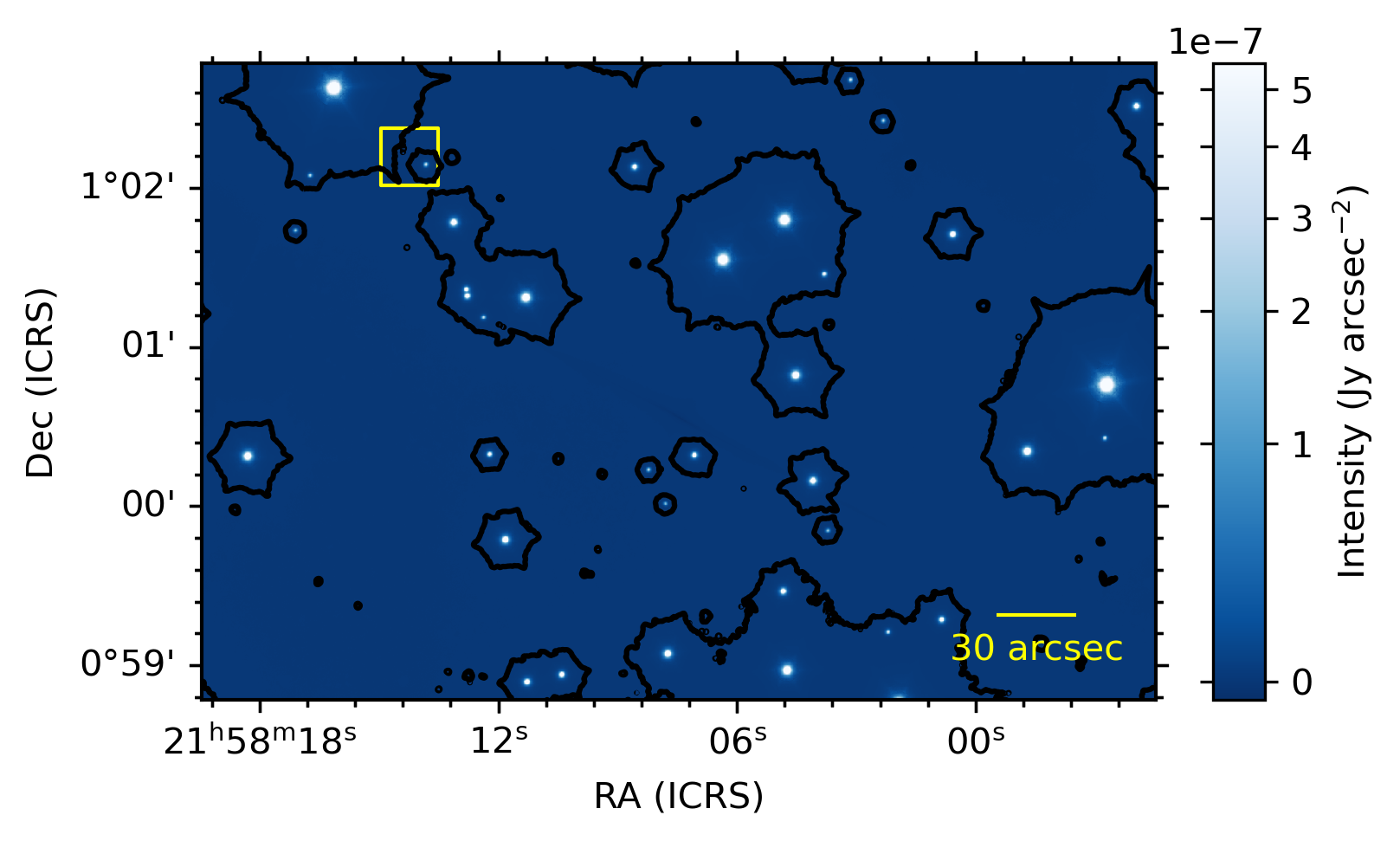}
\includegraphics[trim={50 35 0 0}, clip, width=0.5065\textwidth]{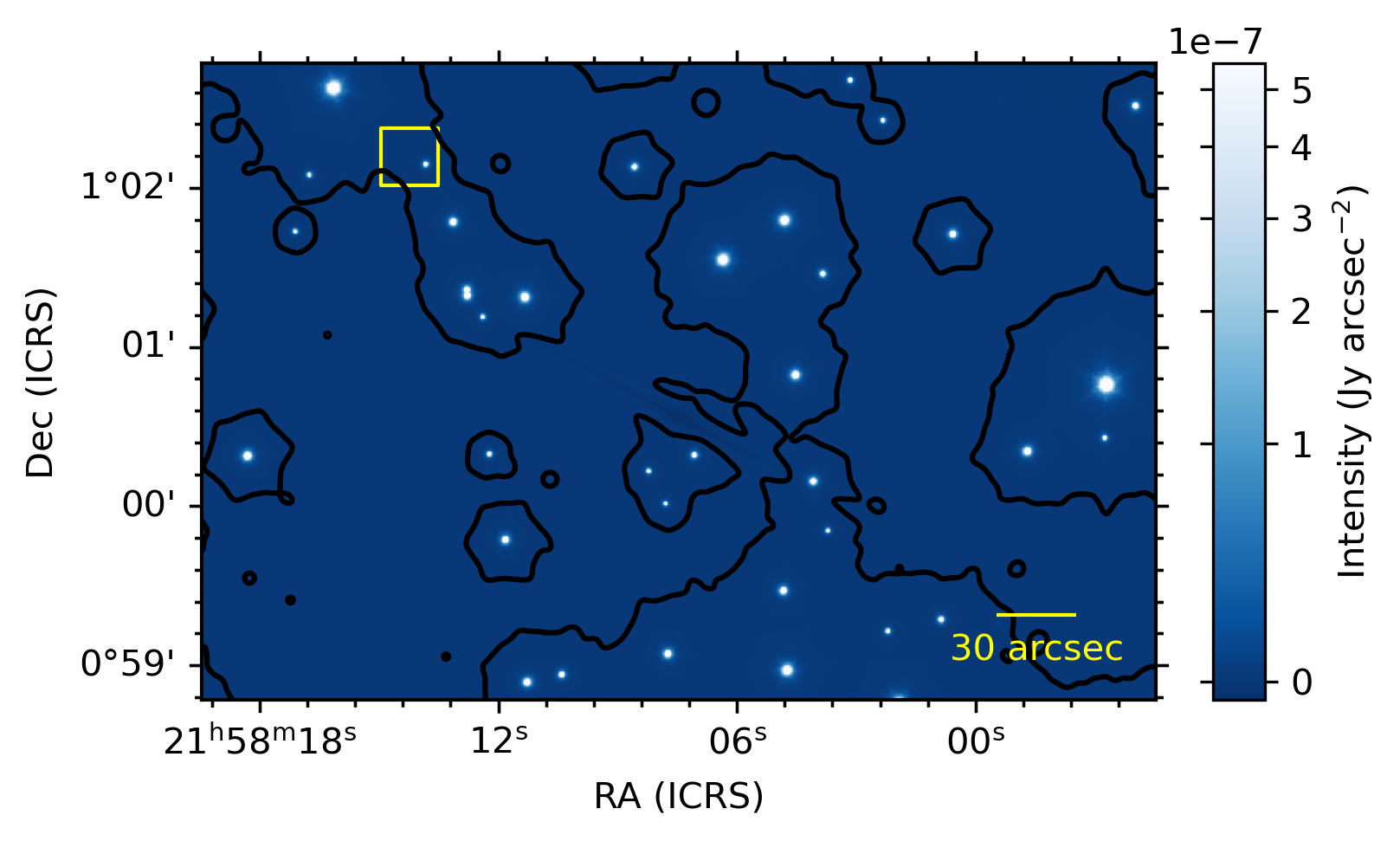}

\includegraphics[trim={0 0 65 0}, clip, width=0.4835\textwidth]{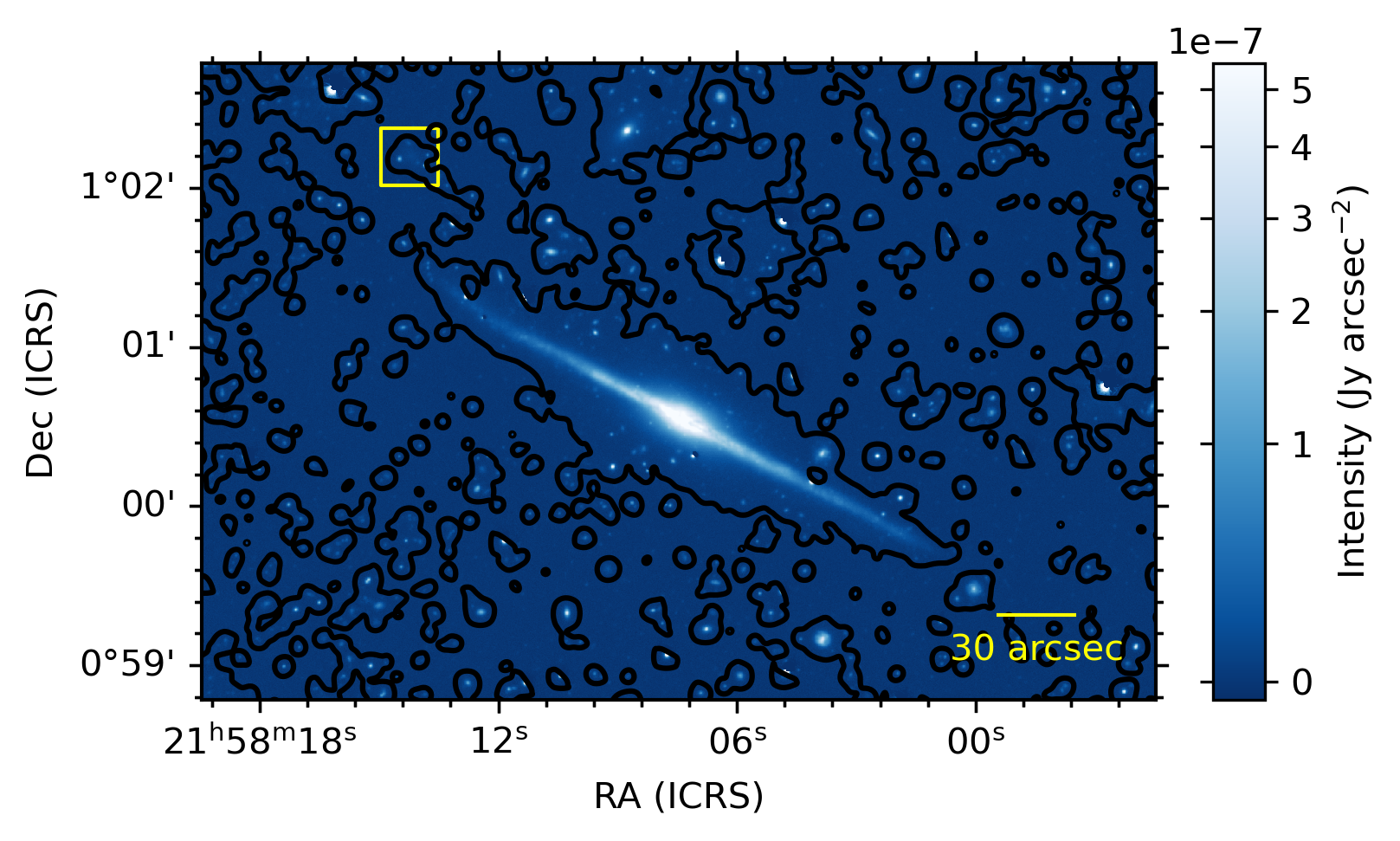}
\includegraphics[trim={50 0 0 0}, clip, width=0.5065\textwidth]{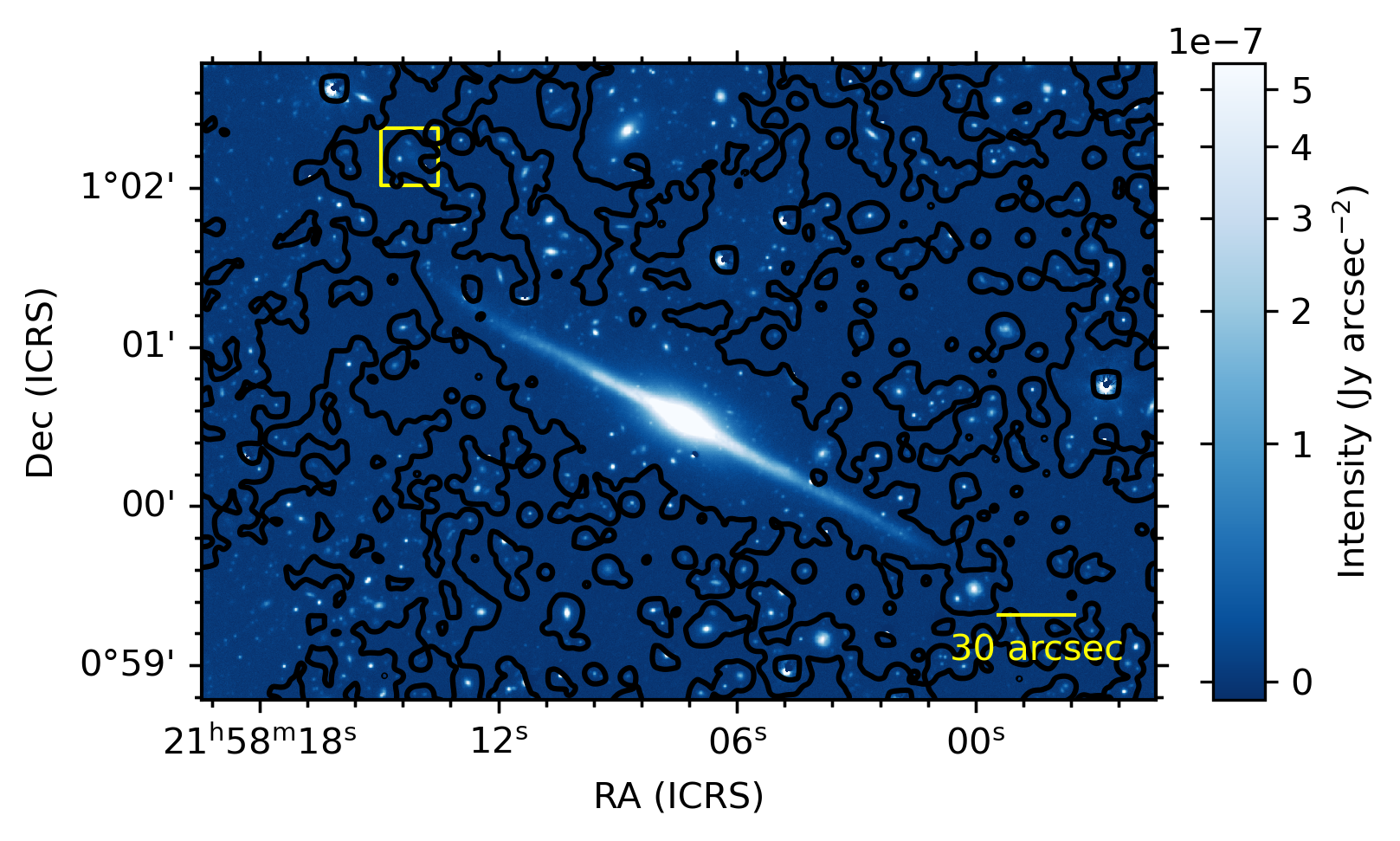}
\caption{Galactic and stellar PSF correction process: \emph{Upper row:} Original mosaics. \emph{Central row:} Model of the foreground stellar emission and galaxy PSF contamination, based on the 2D modeling of the sources and the PSF of the GTC/OSIRIS instrument. \emph{Bottom row:} Final corrected mosaics of \ugc. \emph{Left column:} GTC $g$-band. \emph{Right column:} GTC $r$-band model. The localization of the potential satellite galaxy is indicated by the yellow square in each panel.} 
\label{fig:UGC11895_starfield}
\end{center}
\end{figure*}

The emission of foreground stars present in the FOV of \ugc\ was subtracted by employing a fitting procedure similar to that described \citet{Trujillo2016}, using the OSIRIS/GTC PSF derived from a dedicated observation campaign of $\gamma$ Dra\footnote{A detailed GTC/OSIRIS PSF is publicly available at the GTC webpage: \url{http://www.gtc.iac.es/instruments/osiris/}} ($\alpha=269.1516\degree$, $\delta=51.4889\degree$, $V=2.36$ mag). Fig.\,\ref{fig:UGC11895_starfield} we shows the estimated stellar emission field compared to the original frames and the corrected final mosaics. This PSF was constructed using GTC/OSIRIS $r$ band observations from $\gamma$ Dra and the FOV stars from UGC\,00180 \citep{Trujillo2016}. To avoid systematic biases due to the presence of significant differences between the $g$ and $r$ band PSFs, we construct a new $g$ hybrid PSF \citep[see Sec.\,2.3 in][]{Borlaff2017} using the extended wings of the $\gamma$ Dra~GTC/OSIRIS observations, and the inner core of a PSF constructed with the visible stars of \ugc\ in band $g$. The inner core and the outer wings of the PSF models are combined matching their surface brightness profiles, following the prescriptions from \citet{InfanteSainz2020}. 

Following $g$ and $r$ PSF modeling, we proceeded to identify, fit, and subtract the stray-light from the foreground stars. The \emph{Gaia} archive\footnote{ESA/\emph{Gaia} archive: \url{https://gea.esac.esa.int/archive/}} \citep{Gaia2016, Gaia2021} was used to identify 500 stars at an angular distance lower than $0.16\degree$ from the center of \ugc, which corresponds approximately to the maximum radius covered by our observations with OSIRIS. The \textsf{Imfit}\footnote{\textsf{Imfit} is an GNU open-source project available at \url{https://www.mpe.mpg.de/~erwin/code/imfit/} and \url{https://github.
com/perwin/imfit/}} \citep{Erwin2015} \texttt{PointSourceRot} function was used to rotate and scale the PSF models to the flux and position angle of all the identified stars in each exposure. To avoid unwanted systematic effects by saturation of the core of the brightest stars, we masked heavily saturated regions using the masks generated in Sec.\,\ref{Subsec:methods_reduction}.

\subsection{Galaxy PSF correction}
\label{Subsec:methods_galaxy_psf}

In addition to the stray-light emission from nearby stars, we must take into account the PSF dispersed light from the galaxy itself. This important systematic effect could generate artificial halos and flares in the outskirts of galaxies \citep{Sandin2014, Sandin2015}, distort the vertical structure of edge-on galaxies \citep{martinezlombilla+2019aap629_12}, generate artificial up-bending surface brightness profiles, and make stellar light truncations in the outskirts of the disks undetectable \citep{Borlaff2017}. To avoid a false positive on extended structures in the outskirts of \ugc, a two-dimensional model-based deconvolution method \citep{Trujillo2013} is employed, as summarized below. 
The goal is to generate a model that, when convolved with the PSF, fits the two-dimensional emission of the galaxy. The model must be as accurate as possible, avoiding over-fitting or over-subtraction of any regions of the galaxy. This model is generated using \textsf{Imfit} by starting with a simple model of an edge-on disk using the \texttt{BrokenExponentialDisk3D} function and then building up complexity that provides an optimized fit having minimized amplitude and structure of the residuals. We found that two \texttt{BrokenExponentialDisk3D} functions (for the thin and thick disk of  \ugc, respectively) provide a good fit for the general structure of the galaxy, leaving only irregular shaped residuals that are mainly caused by the presence of dust lanes along the equator of the disk. 

Once the model is generated, an additional model is generated without PSF convolution. This model is a first order approximation to the morphology of the galaxy without dispersion effects. To take into account the emission of structures not included in the model, residuals from the previously-described PSF convolved model are added. The combination of the two frames is the deconvolved image of the galaxy. We repeat the process for both $g$ and $r$ filters. The impact of this correction is visible in the central panels of Fig.\,\ref{fig:UGC11895_starfield} as a thin stripe of negative counts, corresponding to the corrected and deconvolved emission of the galactic plane. For more details on the process and efficiency tests, we refer to Secs.\,2.3 and 2.7 from \citet{Borlaff2017}, respectively.

\subsection{Surface brightness profiles}
\label{Subsec:methods_profiles}

To study the structure of \ugc\ we generate two types of surface brightness profiles: radial and vertical. The methods used to characterize these profiles are described in this section. 

The radial surface brightness profiles of UGC\,11859 were obtained assuming radial symmetry of the galaxy. The position angles (PA) for each band are measured from the generated \textsf{Imfit} models: 61.75$^{\circ}$ and 62.40$^{\circ}$ for the $g$ and $r$ bands, respectively. We used the PA values to rotate and align the images with the horizontal axis (Fig.\,\ref{fig:UGC11895_satelite}). After regridding the images by aligning the major axis of the galaxy to the pixel grid, a long rectangle with a width of 20 pixels ($\sim5$ arcsec) was positioned over the galactic equator in the regridded images. This width was set to select pixels around the maximum surface brightness of the galaxy. 

A predefined set of radial bins are defined for the generation of surface brightness profiles. To maximize the signal-to-noise (SNR) of the outer bins, logarithmic spacing \citep{Erwin2008} is used. The average and dispersion of the intensity and surface brightness inside each radial bin are computed by combining the pixels of the left and right sides of the galaxy. We use random re-sampling with replacement (bootstrapping) to determine the median and uncertainties ($1\sigma$) of each radial bin. 

To measure the scale length of the galactic disk (\hr) as a function of galactocentric radius, we fit the surface brightness profiles ($\mu(r)$) to the analytical expression of the exponential disk profile \citep{Freeman1970}:  
\begin{equation} \label{eq:radial_exp}
\mu(r) = 
\muzero + \cfrac{2.5}{\ln(10)}\cdot\cfrac{r}{\hr}, \vspace{0.25cm}
\end{equation}

where \muzero\, is the central surface brightness of the disk. In order to detect breaks in the shape of the radial surface brightness profile, this exponential disk fit can be estimated in different ($i=1...n$) sections of the galaxy. We note the different central surface brightnesses and scale lengths as \mui\, and \hr$_\mathrm{,i}$.

To obtain the vertical brightness profiles, we divided the galaxy into vertical slices. Each slice is then divided into bins to study the variation of surface brightness as a function of the distance from the galactic plane ($z$). As with the radial profiles, vertical profiles are constructed with the presumption of the galaxy following vertical symmetry. That assumption is found to not be strictly valid at all radii due to the presence of a warp in one side of the galaxy (see Sec.\,\ref{Subsec:results_warp}). However, in Appendix~\ref{A1:flare_sides} we analyze the results of this section for each side of the galaxy independently, confirming that this approximation of the vertical emission averaged over the disk's radial extent does not bias our results. We provide further analysis on the symmetry of the disk in Secs.\,\ref{Subsec:results_flare} and \ref{Subsec:results_warp}. In this case, we can assume that the pixel samples from equivalent radius and $z$ come from the same parent statistical sample, and combine the four equivalent quadrants.

Similarly to the radial exponential profile, we can assume a vertical exponential variation of the surface brightness with the distance from the galactic equator ($z$):

\begin{equation} \label{eq:mu(z)}
 \mu(z)= \mu(0) + \cfrac{2.5}{\ln(10)}\cdot\cfrac{z}{\hz},
\end{equation}

where $\mu(z)$ is the surface brightness at certain height ($z$) from the center of the galaxy and $\hz$ the scale-height. By fitting this expression to the observed variation of the surface brightness profiles with $z$, $\hz$ can be calculated as a function of the galactocentric radius $r$.


\section{Results}
\label{Sec:Results}

In this section we describe the results of the analysis. Section \ref{Subsec:results_radial_profile} describe the detection of a truncation in the outskirts of UGC\,11859. Sections \ref{Subsec:results_warp} and \ref{Subsec:results_flare} describe the results from the vertical analysis of the disk, and the detection of a warp and a disk flare. In Sec.\,\ref{Subsec:results_satellite} we identify a close low surface brightness galaxy, and analyze the possibility of gravitational interaction as a mechanism for the formation of the previously detected features.

\begin{figure*}[]
 \begin{center}

\includegraphics[trim={40 30 40 0}, clip, width=0.99\textwidth]{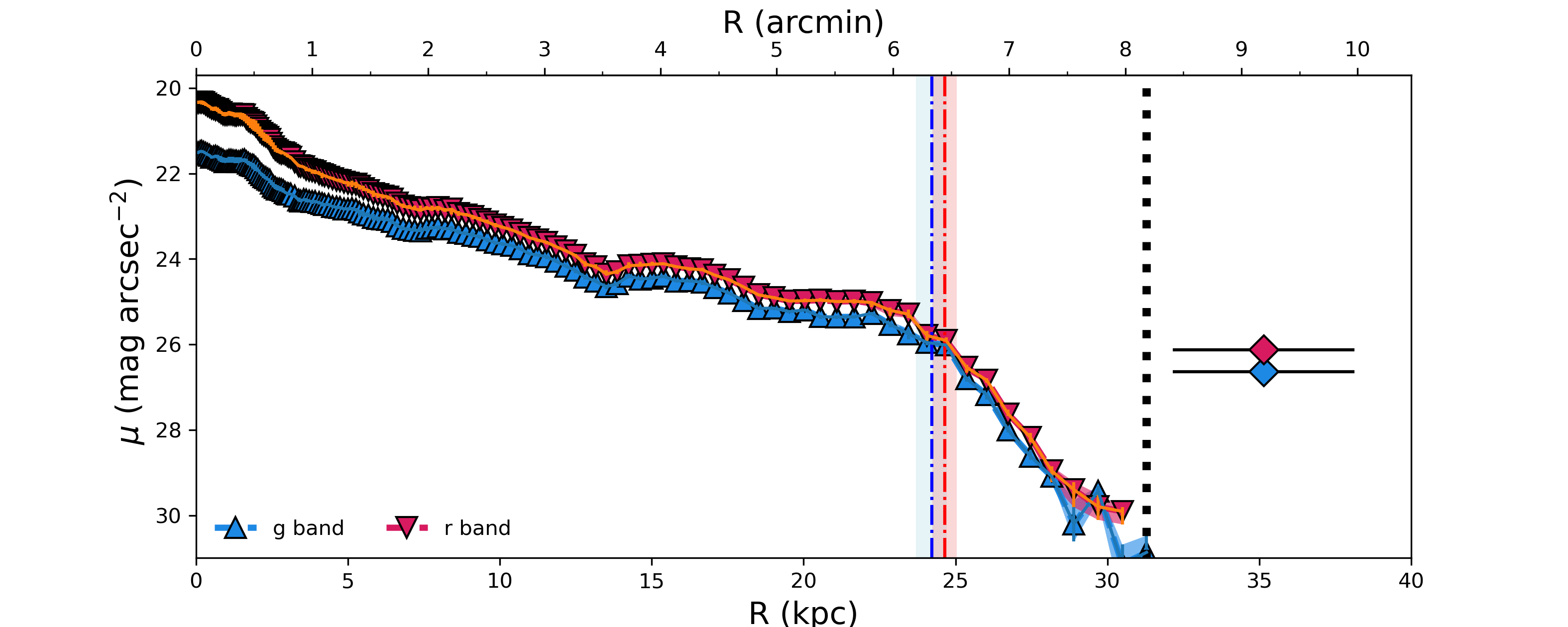}
\includegraphics[trim={40 30 40 30}, clip, width=0.99\textwidth]{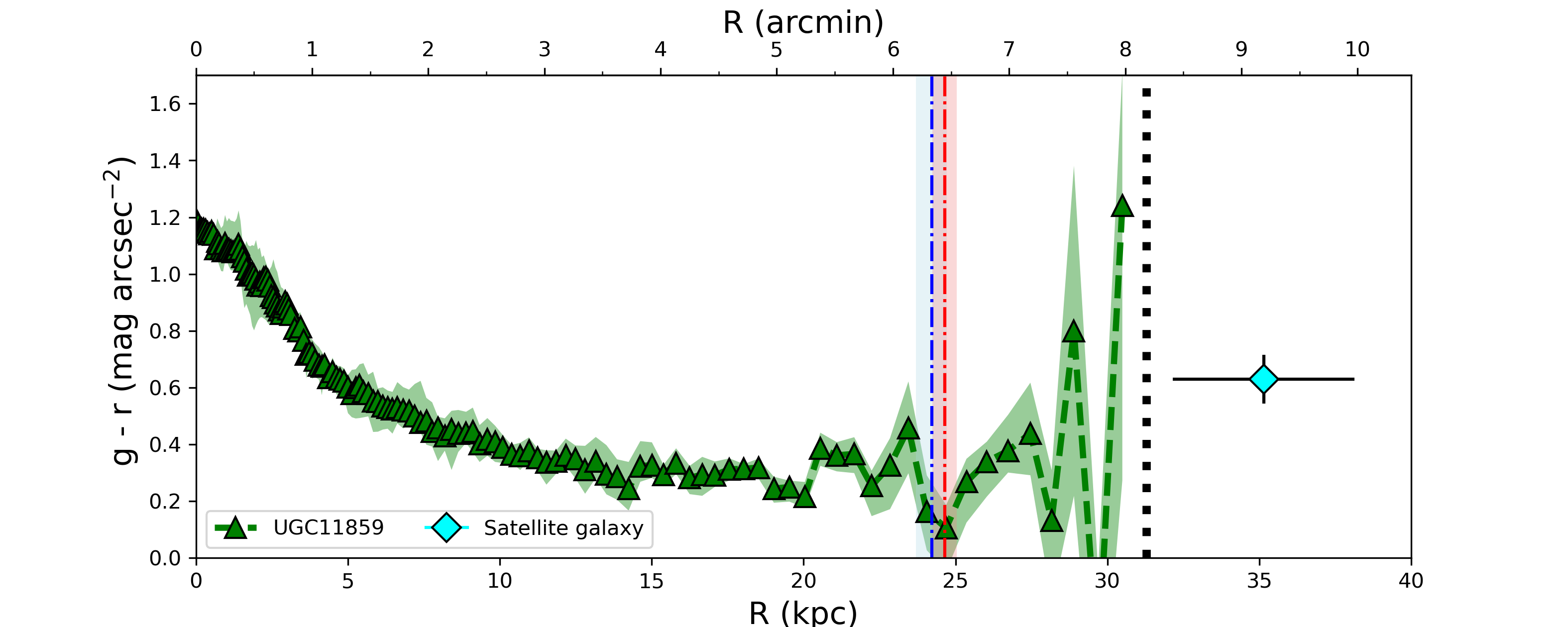}
\includegraphics[trim={40 0 40 30}, clip, width=0.99\textwidth]{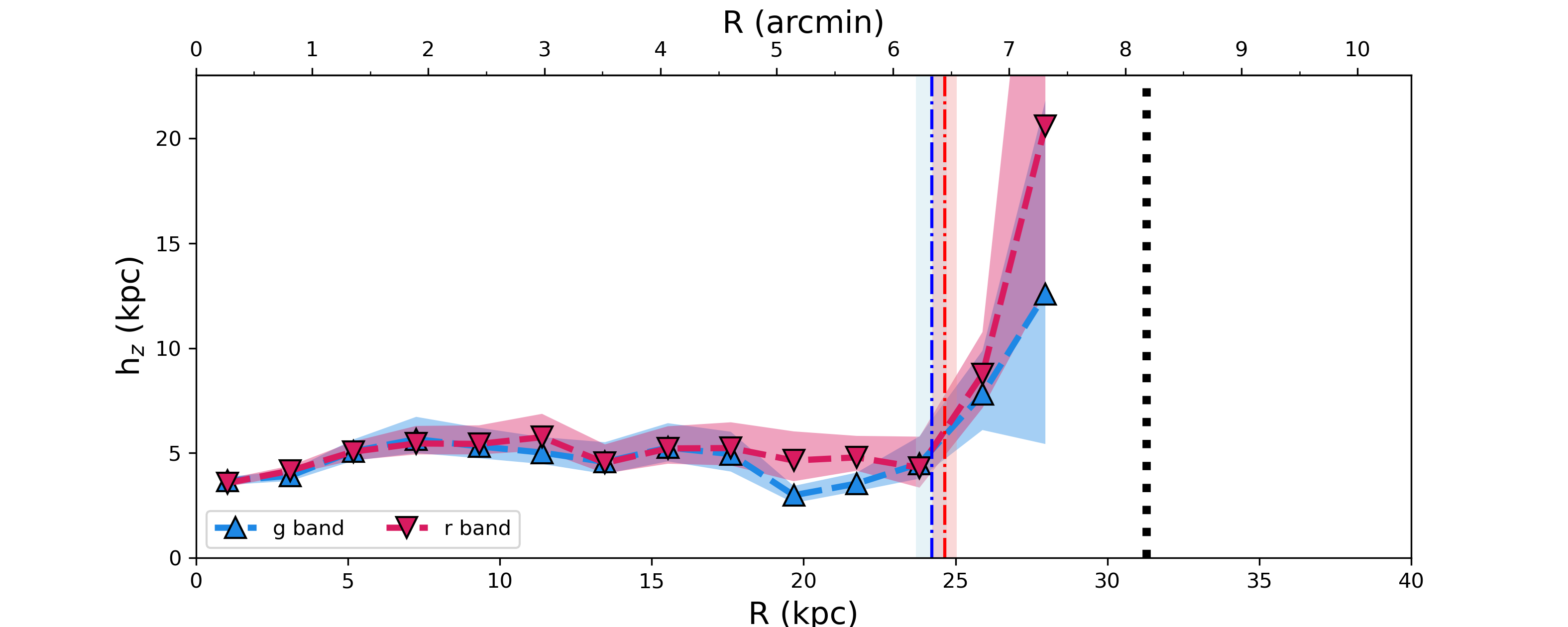}

\caption{\emph{Top:} Radial surface brightness profiles of \ugc\ for \emph{g} and \emph{r} bands (blue triangles and red inverted triangles, respectively). Red and blue diamonds represent the average surface brightness of the potential satellite galaxy GTC-1  found in Sec.\,\ref{Subsec:results_satellite}. The data points in the radial profiles are logarithmically spaced in galactocentric radius. The profiles show an obvious break starting at approximately 24 kpc from the galactic center. \emph{Middle:} Radial $g-r$ color profile of \ugc. The plot shows radial variation of the $g-r$ color and uncertainties of \ugc\ (\emph{green} points and polygons respectively) along the galactic radius. \emph{Teal diamond:} Average color and uncertainty of GTC-1.
\emph{Bottom:} Vertical disk scale-height (\hz) vs. galactocentric radius (R) profile. Data are shown for the bands $g$ (\emph{blue} triangles) and $r$ (\emph{red} inverted triangles) associated with errors (\emph{blue} and \emph{red} shaded areas, respectively). The increase in \hz\ around radius = 24 kpc indicates the presence of a \emph{flare} in the disk of \ugc, associated with the edge-on disk truncation visible in the upper panel.}
\label{fig:UGC11895_radialprofiles}
\end{center}
\end{figure*}

\subsection{Radial surface brightness profile}
\label{Subsec:results_radial_profile}

Fig.\,\ref{fig:UGC11895_radialprofiles} presents the $g$ and $r$ bands radial surface brightness profiles of UGC\,11859 that follow the classic shape of a galactic bright bulge in the inner regions (radius $<5$~kpc) with a central surface brightnesses of $\mug =21.52^{+0.01}_{-0.02}$ \magarc\  and $\mur = 20.32^{+0.02}_{-0.03}$ \magarc\ in the $g$ and $r$ bands, respectively. Beyond that radius, the disk presents a relatively smooth exponential profile that reaches $\sim24$ kpc in radius, where the most notable feature of the surface brightness profile is observed: a clear down-bending break that follows the shape of a edge-on truncation. This break appears in both bands ($g$ and $r$, Fig. \ref{fig:UGC11895_radialprofiles}), indicating that \ugc\ has a "classic" Type~II profile.

We examine the significance and parameters of the different sections of the break profile using {\tt{Elbow}}\footnote[2]{{\tt{Elbow}}: a statistically robust method to fit and classify the surface brightness profiles. The code is publicly available at GitHub (\url{https://github.com/Borlaff/Elbow})} \citep{Borlaff2017}, obtaining a break radius in the $g$-band of $R_{\rm{break}}=1.62^{+0.03}_{-0.03}$ arcmin ($24.2^{+0.5}_{-0.5}$~kpc), a value that is statistically significant at a level of $p=1.2\cdot10^{-3}$. The surface brightness of the break is $\mubreak=25.96^{+0.15}_{-0.06}$~\magarc\ in the $g$-band and $\mubreak=25.91^{+0.38}_{-0.11}$ \magarc\ in the $r$-band. The location of the break is compatible in the observations obtained in the $r$-band, with $R_{\rm{break}}=1.65^{+0.03}_{-0.03}$ arcmin ($23.92^{+0.32}_{-0.07}$ kpc, $p=6.7\cdot10^{-3}$) for the break radius.


\subsection{Vertical profiles of \ugc}
\label{Subsec:results_flare}

Following the method described in Sec.\,\ref{Subsec:methods_profiles}, vertical surface brightness profiles were made to obtain scale-height values \hz\  at certain radial distance. The bottom panel of Fig.\,\ref{fig:UGC11895_radialprofiles} shows the plot of \hz\ vs galactocentric radius (in kpc) and associated errors. Our analysis shows that at $\sim24$ kpc, both $g$ and $r$ bands trace a sudden increase in the scale-height of the galactic disk of \ugc, from an average value of $<h_{z,g}>=4.53^{+0.58}_{-0.41}$ kpc in the band $g$ and $<h_{z,r}>=4.95^{+0.69}_{-0.59}$ kpc in the band $r$ to a peak of $<h_{z,g}>=11.7^{+6.3}_{-6.2}$ kpc and $<h_{z,r}>=19.0^{+14.7}_{-6.5}$ kpc in the $g$ and $r$ bands respectively (thus, a statistically significant change of the \hz\ in both bands).

Interestingly, the increase in \hz\ values occurs:
\begin{enumerate}
    \item in both $g$ and $r$ bands,
    \item in both sides (East and West) of the galaxy (see Fig.\,\ref{fig:flare_sides_galaxy}, in Appendix \ref{A1:flare_sides}),
    \item in all cases, beyond the $\sim24$ kpc galactocentric limit, associated with the edge-on truncation of the galaxy radial light profile.
\end{enumerate}

These attributes of the vertical profile feature suggest that the break could be related to the presence of a \emph{flare} in the disk of \ugc. A sudden increase on the \hz\ of the galactic disk could reduce the apparent surface brightness on the external regions to the flare, and reduce drastically the surface brightness profile as observed (Fig.\,\ref{fig:UGC11895_radialprofiles}). These results suggest that this could be the first observational support for a connection between truncations and flares, a model first explored in \citet{Pohlen2007} and more recently by \citet{Borlaff2016}. In addition, the presence of a flare in both sides of the galaxy  confirms that it is not an artifact caused by the warp.

\subsection{Detection of a warp in the galactic disk}
\label{Subsec:results_warp}

\begin{figure*}[]
 \begin{center}
\includegraphics[trim={80 50 40 55}, clip, width=1.0\textwidth]{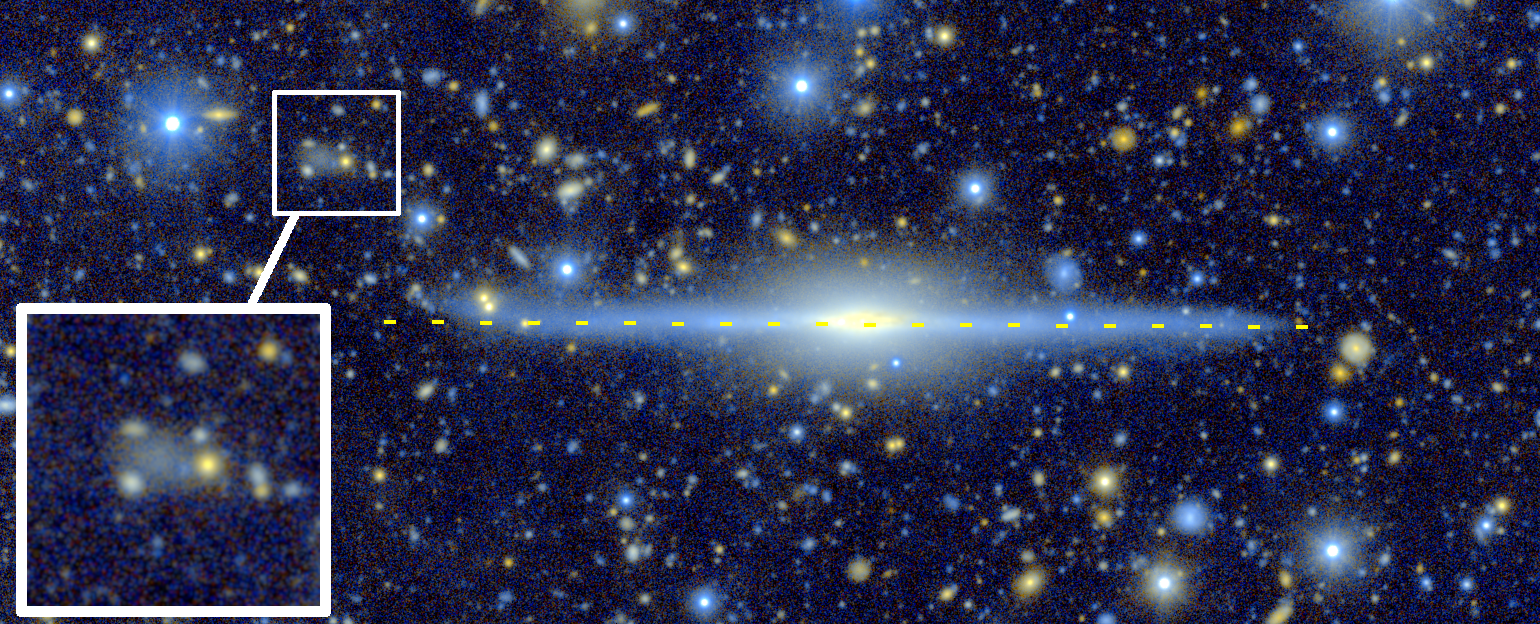}

\caption{Luminance-RGB image aligned to the major axis of \ugc, showing the galactic plane with a yellow dashed line. A misalignment of the galactic disk with respect to the average equator is detectable on the extreme left side of the galaxy. The color information allows us to diagnose that the apparent misalignment cannot be associated with the effect of foreground stars or other background objects.}
\label{fig:lrgb_warp}
\end{center}
\end{figure*}

\begin{figure*}[]
 \begin{center}

\includegraphics[trim={0 0 0 0}, clip, width=1.0\textwidth]{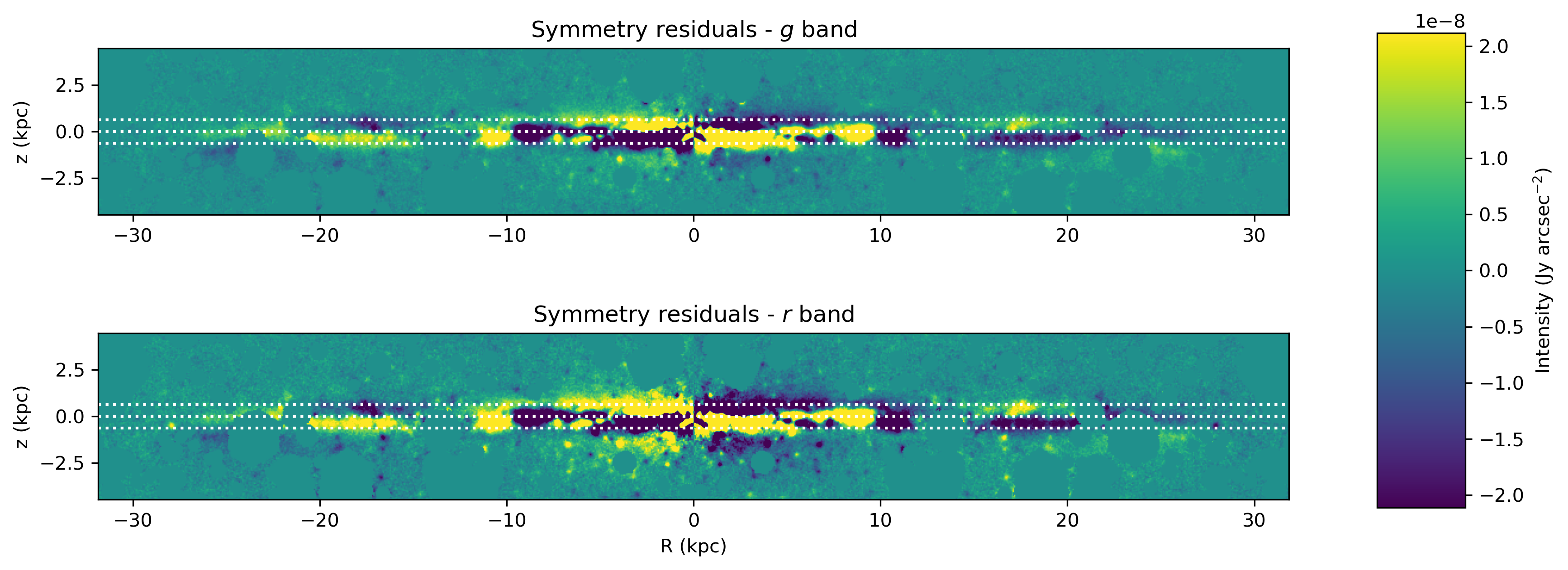}

\caption{Symmetry analysis for \ugc. Top and bottom panels represent the residuals of the $g$ and $r$ band images of \ugc\ after subtraction of the same image but inverting the horizontal ($x$) axis. The horizontal lines represent the upper, equator, and lower limits of the slit used to generate the residual profiles from Fig.\,\ref{fig:symmetry_profiles}. Positive and negative regions dominate across opposite intervals above and below this line, showing the clear sign of a warp in the galactic disk.}
\label{fig:symmetry_images}
\end{center}
\end{figure*}

An interesting result is the presence of a significant warp in the disk of \ugc\ which appears at a radius where the surface brightness has fallen to low values. In Fig.\,\ref{fig:lrgb_warp} we show that near the outer limit of one side of the disk (East side direction) the plane of the galaxy is not symmetric about the galactic plane (reference yellow dashed thick line). Interestingly, this asymmetry is not clearly present on the other side of the disk (West side).
To quantify the asymmetry of the galaxy, we flip across the minor axis and subtract one half of the galaxy from the other, in both the $g$ and $r$ images (East side subtracting West), leaving the vertical orientation untouched. The result of this process is shown in Fig.\,\ref{fig:symmetry_images}. The residual $g$ and $r$ bands show a characteristic pattern: negative residuals dominate the $z>0$ (above the galactic plane) and $r>0$ kpc region up to $r\sim10$ kpc, where the trend inverts. The opposite behaviour can be observed in the region with $z<0$ (below the galactic plane). This change is indicative of a warped (S-shape) galaxy disk and is impossible to reproduce with a perfect (non-warped) disk model, even if the position angle is misaligned. 

\begin{figure*}[]
 \begin{center}
\includegraphics[trim={0 0 0 0}, clip, width=1.0\textwidth]{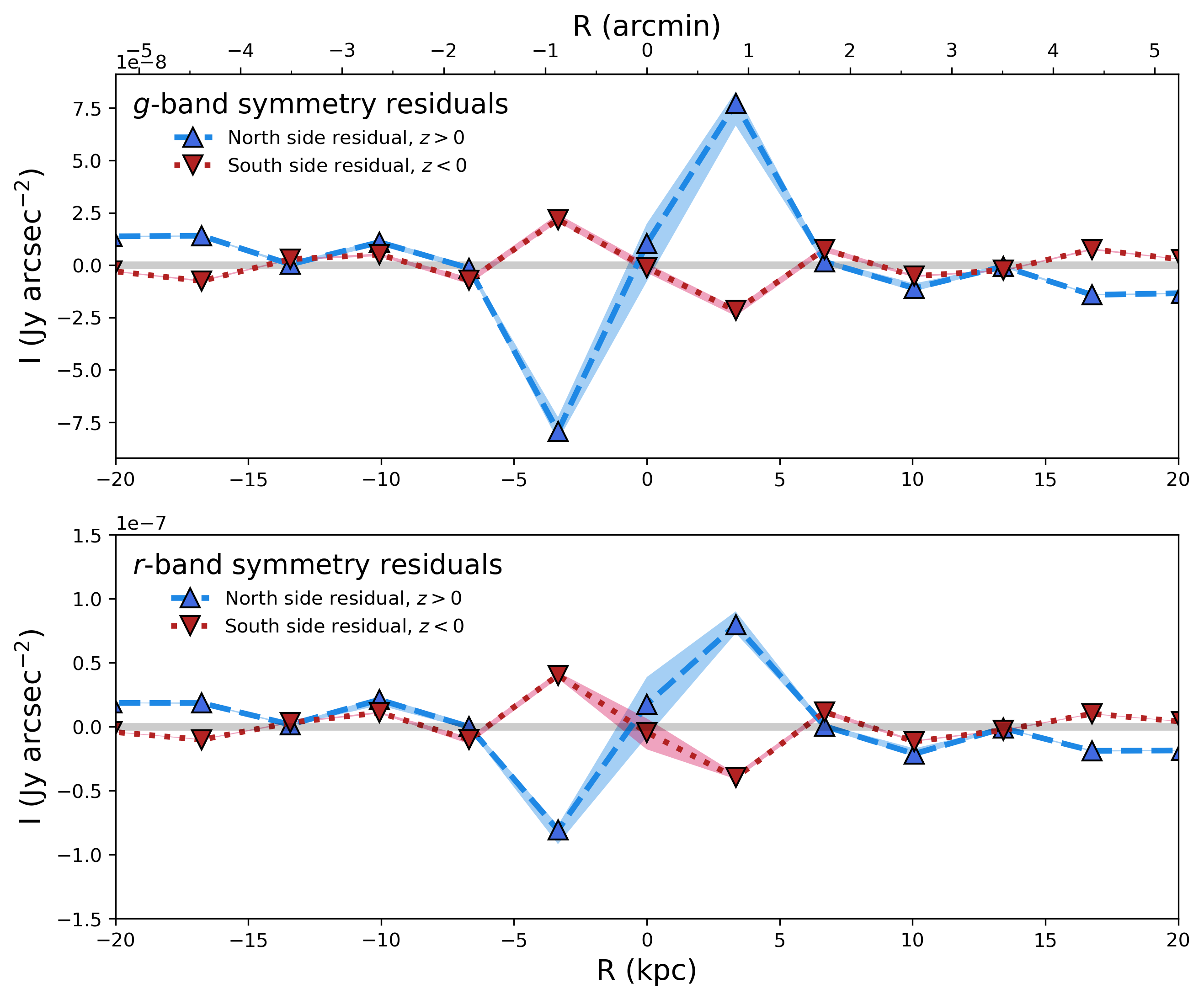}

\caption{Profile of symmetry residuals for \ugc. Top and bottom panels represent the average residuals of the $g$ and $r$ band images of \ugc\ as a function of the galactocentric distance over ($z>0$, blue upward triangles) and under ($z<0$, red downward triangles) the galactic plane. The horizontal solid gray line represents the galactic equator.}

\label{fig:symmetry_profiles}
\end{center}
\end{figure*}

To further quantify the significance of this change in the sign of asymmetries as a function of the galactocentric radius, we generate radial profiles for the regions above and below the galactic plane (Fig.\,\ref{fig:symmetry_profiles}). The results for $g$ and $r$ bands are shown in the top and bottom panels, respectively. In both bands, the residuals maps show the same trend of positive and negative residuals, significantly different from zero, and complementary for the regions above and below the galactic plane, confirming an S-shape in the major axis that could only be generated by the presence of a warp.

\subsection{Identification of a potential satellite galaxy}
\label{Subsec:results_satellite}

\begin{figure*}[]
 \begin{center}
 
\includegraphics[trim={0 0 0 0}, clip, width=0.618\textwidth]{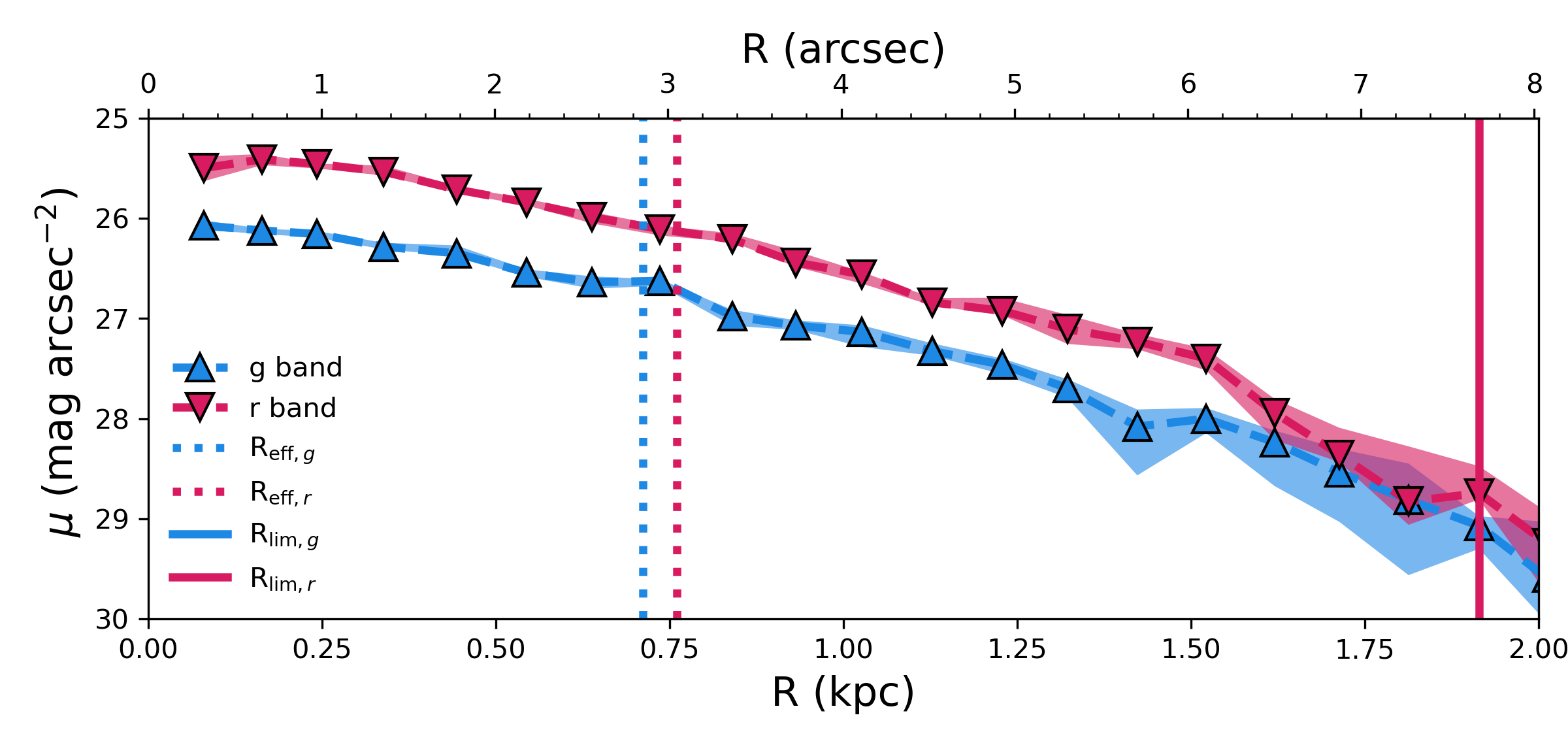}
\includegraphics[trim={0 0 0 0}, clip, width=0.377\textwidth]{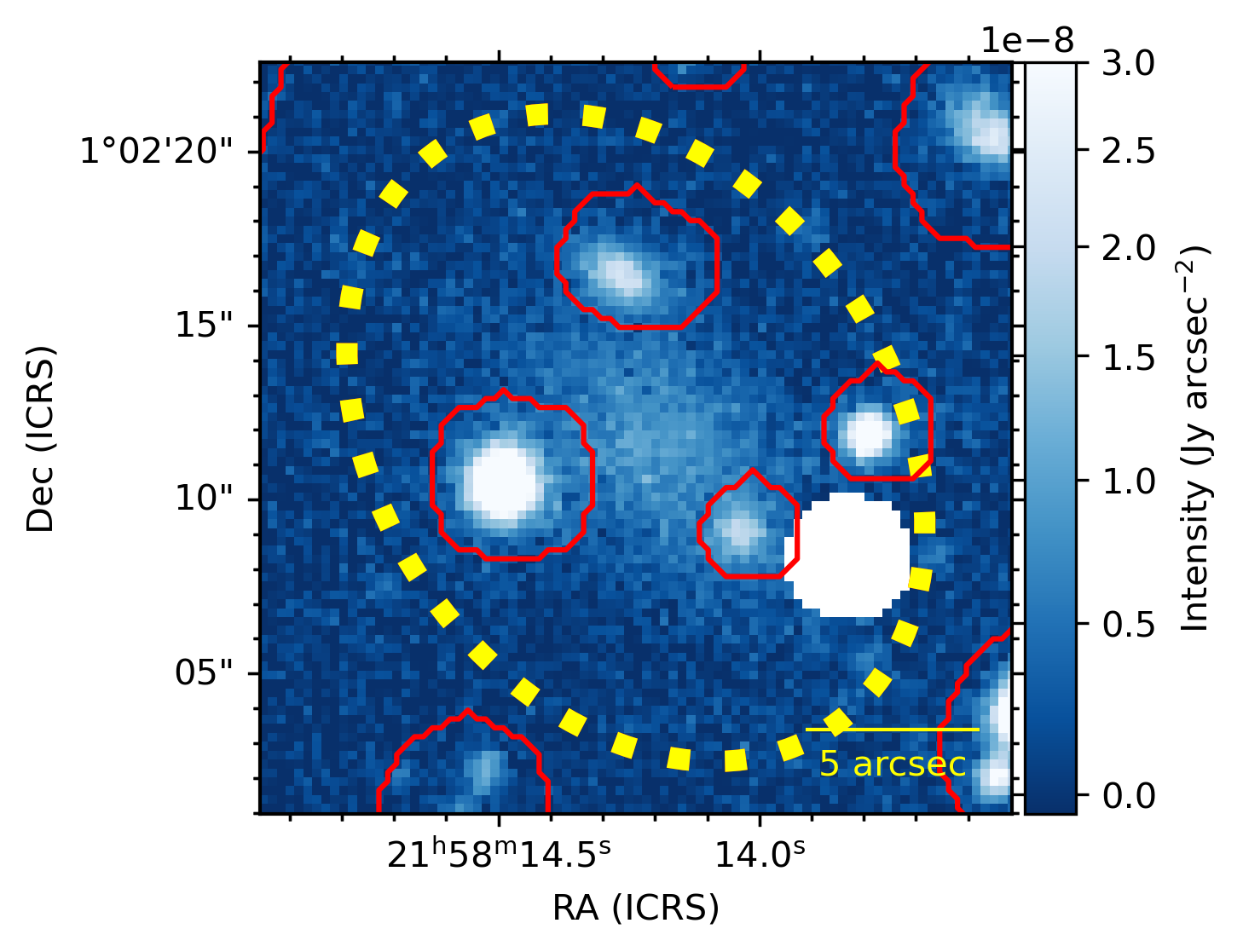}

\includegraphics[trim={0 0 0 25}, clip, width=0.618\textwidth]{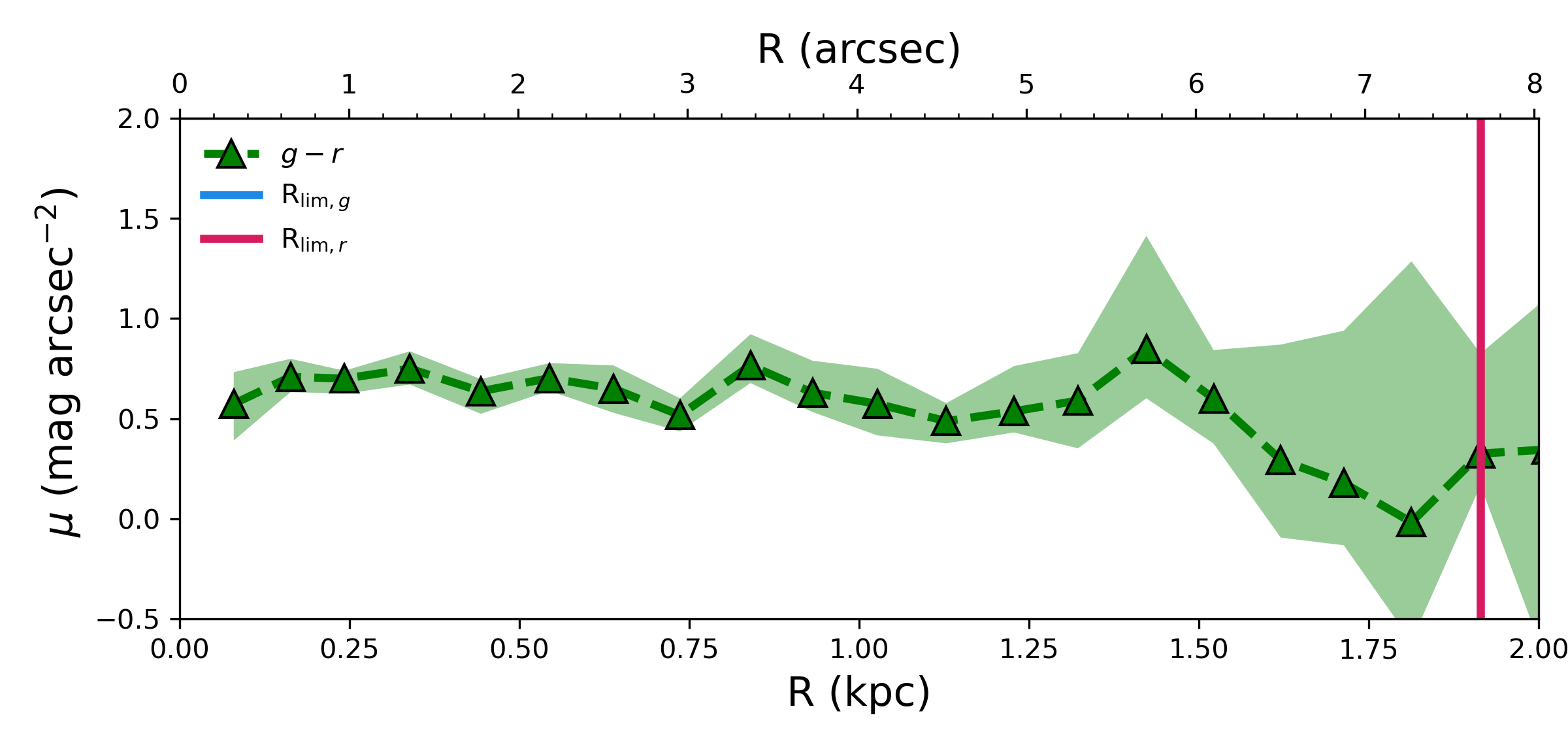}
\includegraphics[trim={0 0 0 0}, clip, width=0.377\textwidth]{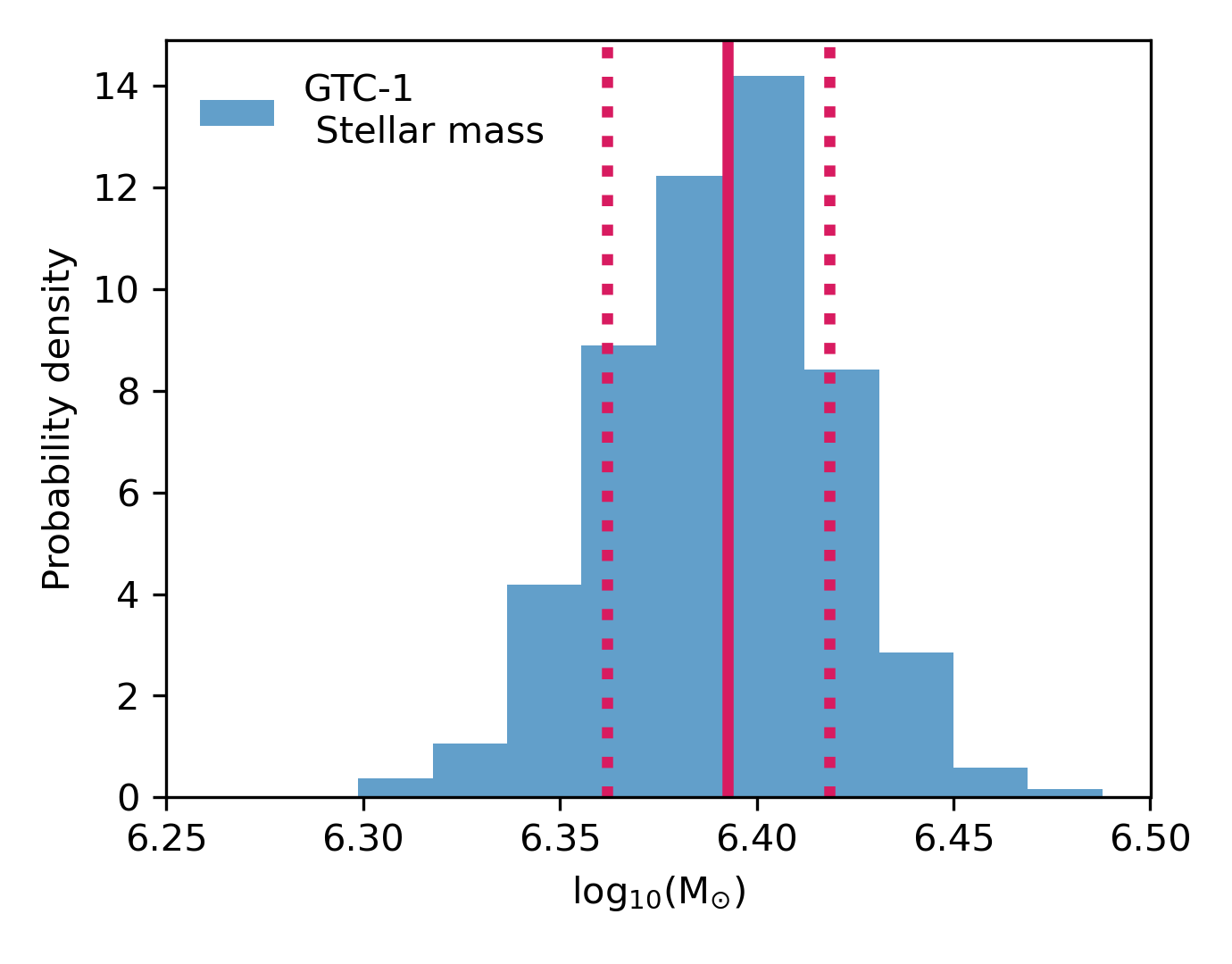}

\caption{Results from the analysis of the potential satellite galaxy GTC-1. \emph{Upper and Lower-left}: Surface brightness profile and \emph{g-r} color profile using the $g$ and $r$ GTC observations. Dashed lines represent the effective radius ($R_{\rm{eff}}$) in each band, and the solid vertical lines the limiting radius ($3\sigma$). \emph{Upper-right}: Image of GTC-1 showing intensity in $g$ band. Red contours corresponded to masked regions generated for stellar mass and radial profile estimations. Yellow dashed ellipse represent the limit radius for detection in $g$ band. \emph{Lower-right}: Probability density of the stellar mass of GTC-1, following the recipes from \citet{Bakos2008} and assuming that it is a satellite of UGC11859.}
\label{fig:ugc11859_df1}
\end{center}
\end{figure*}

Located close to the edge of the disk in the northeast region of \ugc\ we found a diffuse object previously not detected in SDSS and \emph{Gaia} data ($\alpha=329.5591^\circ$, $\delta=+1.0367^\circ$). The new object, which we identified as 21h58m07.33s/+01°00'31.95".Grantecan 1 and we will refer tentatively as GTC-1 due to its apparent -- although not yet confirmed --  association, shows a diffuse, non-point source morphology, eliminating the possibility of it being a faint star. The central surface brightness is particularly low $\mu_{g} =  26.06^{+0.04}_{-0.08}$~mag~arcsec$^{-2}$ ($g$ band) and $\mu_{r} = 25.50^{+0.15}_{-0.13}$~mag~arcsec$^{-2}$ ($r$ band), being similar to low surface brightness galaxies \citep[$\mu_0 > 24$ \magarc\ in the SDSS $g$ band, see][]{Roman2021A&A656A44R}. The very low surface brightness of GTC-1 explains its absence in the catalogs from previous surveys (SDSS), but the object is detectable in the Stripe 82 mosaics as well \citep{IACStripe82}. The surface brightness profile of the galaxy is shallow, without a visible bulge at its center (Fig.\,\ref{fig:ugc11859_df1} upper left panel). The effective radius ($R_{\rm{eff}}$) of the galaxy is 2.8--3.0 arcsec, with a surface brightness at the $\mu_{\rm{eff}, g} = 26.63^{+0.05}_{-0.04}$~mag~arcsec$^{-2}$ ($g$ band) and $\mu_{\rm{eff}, r} = 26.13^{+0.03}_{-0.07}$~mag~arcsec$^{-2}$ ($r$ band), and it can be detected up to a limiting radius of 10.0 and 7.7 arcsec ($g$ and $r$ band, respectively).

GTC-1 is located at an angular distance of $\sim45$ arcsec from the detectable edge of \ugc, in the same direction of the distortion that we identified as a galactic warp. The average color ($g-r$) of this object is $0.63^{+0.05}_{-0.07}$ mag (Fig.\,\ref{fig:ugc11859_df1} bottom left panel), a value significantly redder than the extremities of the main galaxy \ugc, but compatible with that of \ugc's inner disk (see Fig.\,\ref{fig:UGC11895_radialprofiles}). Under the assumption that GTC-1 and \ugc\ have similar redshifts ($D=51.4$~Mpc for \ugc, from \citet{2010Ap&SS.325..163P}), GTC-1 would have a projected distance of $\sim35$~kpc from the center of \ugc. Following the recipes from \citet{Bakos2008} to estimate the surface stellar mass density from SDSS $g$ and $r$ photometry, we compute the total stellar mass of GTC-1 (see right panel of Fig.\,\ref{fig:UGC11895_satelite}) to be $\log_{10}$($M_\odot$) $=6.39^{+0.03}_{-0.03}$ (Fig.\,\ref{fig:ugc11859_df1} bottom right panel).

The apparent proximity, and the coincident direction of the warp in \ugc\ suggest that this object could be a satellite galaxy of \ugc. This supports the existence of a population of satellite galaxies at very low surface brightness, that could be formed by, or participating in the formation of the outskirts of larger, galaxies. We provide a detailed discussion in Sec.\,\ref{Sec:Discussion} of the compatibility of the observations of GTC-1 in the context of the current knowledge of dwarf and ultra-diffuse galaxies.

\section{Discussion}
\label{Sec:Discussion}

The two-dimensional photometric analysis of \ugc\ provides interesting insights regarding the three-dimensional structure of the galaxy. Disk warps and flares are very common galaxy morphology features in HI and the evidence from \ugc\ suggests that there might well be coincidence between the gas structure and the star structure in the outer galactic regions, which would support the hypothesis that stellar warps and flares are formed within gas warps and flares \citep{vanderKruit2007A&A466883V}. More recent work by \citet{MartinNavarro2012} combined optical images from the Sloan Digital Sky Survey (SDSS) and near infrared images from the Spitzer Survey of Stellar Structure in Galaxies (S4G) to produce radial profiles of 34 near-edge-on galaxies. Their analysis revealed Type~II breaks at an average radius of $\sim$8 kpc and a second, sharper truncation at $\sim$14 kpc, concluding that two different physical mechanisms were responsible for the two breaks. Breaks were associated with thresholds in the star formation efficiency and the sharper truncations with the location of maximum angular momentum of the material forming the galaxies. These results imply that the truncations detected at earlier epochs by \citet{vanderkruit1979, vanderKruit1981a, vanderKruit1981b} correspond to the foreshortened breaks according to \citet{MartinNavarro2012}. In contrast, the models presented in \citet{Borlaff2016} suggests that down-bending profiles can be also be a projection artifact if galactic flares are both sufficiently strong and observed in edge-on orientation. The results presented in this work, showing a coincidence between the presence of a strong flare and a Type~II break, provide for the first time observational support to this hypothesis. 
 
Warps like the one present in the stellar disk of the Milky Way \citep{Skowron2019Sci...365..478S} have been associated with the gravitational interaction with a misaligned flattened halo \citep{Sparke1988.10.1093mnras234.4.873}, infall of cosmic material into the galactic disk \citep{Shen2006MNRAS.370....2S}, or satellite galaxies \citep{Weinberg2006ApJ...641L..33W, Laporte2019MNRAS.485.3134L,Poggio2020NatAs...4..590P}. Interestingly, all these three mechanisms are associated with the effect of low surface brightness sources (halos, satellites, the cosmic web) on high surface brightness structures. \citet{vanderKruit2007A&A466883V} analyzed the presence of \HI-disk warps and stellar truncations, concluding that truncations and \HI-warps are spatially coincident in their sample, pointing towards an evolutionary scenario where the inner disk forms first and the warped outer disk forms much later due to gas infall of gas with a higher angular momentum in a different orientation. We must note that while similar analyses of stellar disks have been performed before \citep{deGrijs1997}, they have never been performed at the low surface brightness levels achieved in this work. 

The presence of a warp and a flare in the outer disk, coincidentally with a break in the surface brightness profile poses the question of whether the phenomena could be related, although the apparent one-sidedness of the warp but not of the flare suggests otherwise. The precession and winding of warped material in a self-gravitating rotating disk might lead to an increase of the thickness of galactic disks at high radii. For example, in M\,83, \citet{Rogstad1974ApJ...193..309R} estimated a winding time-scale of 1~Gyr. For reference, with a rotation velocity of $V\sim170$ km s$^{-1}$ \citep{Haynes2018}, \ugc\ would take $\sim0.7$~Gyr to spin once. However, mechanisms for producing persistent warps have been suggested \citep{Sparke1988.10.1093mnras234.4.873}. Further spectroscopic observations and kinematic analyses would be required to explore this hypothesis in \ugc.

The shape of the color profile displayed in Fig.\,\ref{fig:UGC11895_radialprofiles} is commonly observed among spiral galaxies \citep[][]{James2008}, with a redder central region ($g-r\sim1.1$) and progressively bluer colors with increasing galactocentric radius out to the edge of the disk ($g-r\sim0.3$).  These blue colors are indicative of actively forming stars, suggesting the availability of a significant reservoir of diffuse gas. The color gradient suggests a significant star formation rate in the galaxy outskirts and possibly a high gas content, consistent with previous estimations of the total \HI\ gas content ($\log_{10}(\Msun_{\rm{, HI}})=9.92\pm0.06$) \citep{Haynes2018ApJ86149H}. This could be an indirect indication of the on-going accretion of pristine material from the cosmic web, one of the suggested mechanisms to form warps. We note that our study is focused on the properties of the outer disk of \ugc\ ($r>25$ kpc), where dust absorption effects are expected to be minimal. In particular, \citep{martinezlombilla+2019aap629_12} describes \ugc\ as a very thin disk galaxy, without signs of strong dust lanes. If strong dust lanes were generating the features observed (flare, break) we would expect a break in the $g-r$ color profile of the main galaxy, which is not observed (see Fig.\,\ref{fig:UGC11895_radialprofiles}). We conclude that the effects of dust are most likely negligible for the study of the outer structure of the galaxy, although higher resolution observations with HST would allow improving this approximation.

Our results suggest the possibility that GTC-1 is part of such infall material in the surroundings of \ugc. The association of GTC-1 to \ugc\ is still pending spectroscopic observations. However, the presence of the possible satellite galaxy near the edge of the galactic disk, the presence of the disk warp in the same direction support the possibility that this satellite could have formed in the environment of the main galaxy and that it shares in the accretion process with the interstellar medium of \ugc. Assuming a distance to GTC-1 equal to that to the main galaxy, the stellar mass is $\log_{10}$($M_\odot$) $=6.39^{+0.03}_{-0.03}$. We can compare GTC-1 with known data from dwarf and ultra-diffuse galaxies(UDGs): Taking into account its effective radius ($R_{\rm{eff}}\sim3"$, $\sim$0.7 kpc), we can estimate an effective surface stellar mass density ($\Sigma_{\rm{eff}}= (1/2)\Msun/R_{\rm{eff}}^{2})$) equal to $\Sigma_{\rm{eff}}=2.4\cdot10^{6} \Msun kpc^{-2}$, falling near the lower end of the range ($0.2\cdot10^{7} \Msun kpc^{-2} < \Sigma_{\rm{eff}} < 2.9\cdot10^{7} \Msun kpc^{-2}$) proposed by \citet{Koda2015ApJ807L2K} and \citet{Sales2020MNRAS4941848S} as a criterion which classifies a galaxy as an UDG. Even though these objects were observed and modeled in galaxy cluster environments, \citet{Sales2020MNRAS4941848S} found that the UDG population contained objects similar to field dwarf galaxies. Also, if we include and compare the average colors of the UDGs, which have values $g-r=0.67$ \citep{Koda2015ApJ807L2K} and $g-r=0.59$ \citep{Venhola2017A&A608A142V},  with those of GTC-1 ($g-r=0.63\pm0.07$), we may suggest that our preliminary results for this object are in line with the relevant published data. 

The galactocentric distance of GTC-1 would be $\sim35$ kpc, a distance similar to that of Segue 1 \citep{Fritz2018ApJ860164F} or the Sagittarius Dwarf Spheroidal Galaxy to the center of the Milky Way \citep{Ibata1995MNRAS277781I}. The gravitational forces that dominate the galactic gaseous disk in the outer regions also shape the structure of the stellar disk. Unfortunately, the latter is usually more difficult to detect observationally. In the present study we have shown that the disk of \ugc\ presents a slight (but significant and detectable) disturbance compatible with the presence of a gravitational interaction, either with the satellite galaxy or with another object. Spectroscopic observations to acquire relative redshifts would confirm the association of this new object with \ugc.


\section{Conclusions}
\label{Sec:Conclusions}

We have obtained deep photometric images in the $g$ and $r$ bands of the edge-on disk galaxy \ugc. After performing very careful reduction procedures we have deduced the following:

\begin{enumerate}
    \item The radial surface brightness profile of the disk shows a clear break and a sharp fall at $\sim$24 kpc galactocentric radius, characteristic of an edge-on truncation. This is observed in both $g$ and $r$ bands.
    
    \item The vertical surface brightness profiles show an increase in scale height \hz\ with increasing galactocentric radius, which is more clearly revealed after applying a stellar pollution correction model, which improves significantly the fit for the $r$ band, showing that the observed variation is not a systematic contamination effect due to scattered light within the image or from external objects. The presence of this flare supports the suggestion that edge-pn disk truncations can be produced by the combination of a flare with the down-bending profile of a conventional Type~II disk \citep{Borlaff2016}. To detect this type of flare requires meticulous photometry down to the detection limits achieved in the present study.
    
    \item The images, both in $g$ and $r$ bands, show the presence of a warp at one side of the stellar disk.
    
    \item Close to this end of the disk we detect a possible small satellite galaxy, which we have tentatively termed GTC-1. Assuming that this object is in fact close to \ugc we have estimated its mass as $\log_{10}$($M_\odot$) $=6.39^{+0.03}_{-0.03}$.
    
    \item The color profile of \ugc  shows that the galaxy is redder in the central zone, tending towards blue in the outer disk, which suggests a significant star formation rate and higher gas content there.
    
\end{enumerate}

Ultra-deep surveys, studying galaxies such \ugc\ presented in this work, should be the basis for future observations of large samples (order of hundreds of nearby galaxies). Such data will allow the description and characterization of the structure and stellar populations comprising the outer, low surface brightness regions of objects from a statistical point of view, and most importantly, to provide a census of the low surface brightness satellites and structures around galaxies in the local universe, something only achieved in the environment of the Milky Way. Robust detection of larger and fainter structures presents technical difficulties. However, it is possible and this motivation should serve as a driver for the improvement and optimization of data reduction methods in future deep surveys such as LSST or \emph{Euclid}.

Finally, we can draw the inference that the detection, using high quality very deep imaging, of a warp and flare in the external disk of \ugc\ opens up the possibility that further analysis of this system will provide insight on the connection between these structures, and for other galaxies, can deepen our understanding of the causes of warps and flares.

\begin{acknowledgements}
L.O. was supported by the National Doctoral Degree Scholarship given by The National Research and Development Agency of Chile (ANID).
A.B. was supported by an appointment to the NASA Postdoctoral Program at the NASA Ames Research Center, administered by Universities Space Research Association under contract with NASA. Support for program AR 17041 was provided by NASA through a grant from the Space Telescope Science Institute, which is operated by the Association of Universities for Research in Astronomy, Inc., under NASA contract NAS 5-03127. A.B. is also supported by a NASA Astrophysics Data Analysis grant (22-ADAP22-0118). This work has made use of data from the European Space Agency (ESA) mission {\it Gaia} (\url{https://www.cosmos.esa.int/gaia}), processed by the {\it Gaia} Data Processing and Analysis Consortium (DPAC, \url{https://www.cosmos.esa.int/web/gaia/dpac/consortium}). This research has made use of the VizieR catalogue access tool, CDS, Strasbourg, France. Funding for the DPAC has been provided by national institutions, in particular the institutions participating in the {\it Gaia} Multilateral Agreement. This research made use of NumPy \citep{harris2020array}, Astropy, a community-developed core Python package for Astronomy \citep{refId0}. All of the figures on this publication were generated using Matplotlib \citep{Hunter:2007}. This work was partly done using GNU Astronomy Utilities (Gnuastro, ascl.net/1801.009) version 0.11.22-dc86. Funding for SDSS-III has been provided by the Alfred P. Sloan Foundation, the Participating Institutions, the National Science Foundation, and the U.S. Department of Energy Office of Science. The SDSS-III web site is http://www.sdss3.org/. SDSS-III is managed by the Astrophysical Research Consortium for the Participating Institutions of the SDSS-III Collaboration including the University of Arizona, the Brazilian Participation Group, Brookhaven National Laboratory, Carnegie Mellon University, University of Florida, the French Participation Group, the German Participation Group, Harvard University, the Instituto de Astrofisica de Canarias, the Michigan State/Notre Dame/JINA Participation Group, Johns Hopkins University, Lawrence Berkeley National Laboratory, Max Planck Institute for Astrophysics, Max Planck Institute for Extraterrestrial Physics, New Mexico State University, New York University, Ohio State University, Pennsylvania State University, University of Portsmouth, Princeton University, the Spanish Participation Group, University of Tokyo, University of Utah, Vanderbilt University, University of Virginia, University of Washington, and Yale University. Finally we would like to thank the anonymous referee for suggestions which have helped us to improve the quality of the article.
\end{acknowledgements}

%
%

\bibliographystyle{aasjournal}
\bibliography{ugc11859.bib}{}

\begin{thebibliography}{}
\expandafter\ifx\csname natexlab\endcsname\relax\def\natexlab#1{#1}\fi
\providecommand{\url}[1]{\href{#1}{#1}}
\providecommand{\dodoi}[1]{doi:~\href{http://doi.org/#1}{\nolinkurl{#1}}}
\providecommand{\doeprint}[1]{\href{http://ascl.net/#1}{\nolinkurl{http://ascl.net/#1}}}
\providecommand{\doarXiv}[1]{\href{https://arxiv.org/abs/#1}{\nolinkurl{https://arxiv.org/abs/#1}}}

\bibitem[{{Akhlaghi}(2019)}]{noisechisel_segment_2019}
{Akhlaghi}, M. 2019, arXiv e-prints, arXiv:1909.11230.
\newblock \doarXiv{1909.11230}

\bibitem[{{Akhlaghi} \& {Ichikawa}(2015)}]{gnuastro}
{Akhlaghi}, M., \& {Ichikawa}, T. 2015, ApJS, 220, 1,
  \dodoi{10.1088/0067-0049/220/1/1}

\bibitem[{{Azzollini} {et~al.}(2008){Azzollini}, {Trujillo}, \&
  {Beckman}}]{Azzollini2008}
{Azzollini}, R., {Trujillo}, I., \& {Beckman}, J.~E. 2008, \apj, 684, 1026,
  \dodoi{10.1086/590142}

\bibitem[{{Bakos} {et~al.}(2008){Bakos}, {Trujillo}, \& {Pohlen}}]{Bakos2008}
{Bakos}, J., {Trujillo}, I., \& {Pohlen}, M. 2008, \apjl, 683, L103,
  \dodoi{10.1086/591671}

\bibitem[{{Battaner} {et~al.}(2002){Battaner}, {Florido}, \&
  {Jim{\'e}nez-Vicente}}]{Battaner2002}
{Battaner}, E., {Florido}, E., \& {Jim{\'e}nez-Vicente}, J. 2002, \aap, 388,
  213, \dodoi{10.1051/0004-6361:20020423}

\bibitem[{{Bertin}(2006)}]{Bertin2006}
{Bertin}, E. 2006, in Astronomical Society of the Pacific Conference Series,
  Vol. 351, Astronomical Data Analysis Software and Systems XV, ed.
  C.~{Gabriel}, C.~{Arviset}, D.~{Ponz}, \& S.~{Enrique}, 112

\bibitem[{{Bertin}(2010)}]{Bertin2010}
{Bertin}, E. 2010, {SWarp: Resampling and Co-adding FITS Images Together}.
\newblock \doeprint{1010.068}

\bibitem[{{Bertin} \& {Arnouts}(1996)}]{Bertin1996}
{Bertin}, E., \& {Arnouts}, S. 1996, \aaps, 117, 393,
  \dodoi{10.1051/aas:1996164}

\bibitem[{{Bertin} {et~al.}(2002){Bertin}, {Mellier}, {Radovich}, {Missonnier},
  {Didelon}, \& {Morin}}]{Bertin2002}
{Bertin}, E., {Mellier}, Y., {Radovich}, M., {et~al.} 2002, in Astronomical
  Society of the Pacific Conference Series, Vol. 281, Astronomical Data
  Analysis Software and Systems XI, ed. D.~A. {Bohlender}, D.~{Durand}, \&
  T.~H. {Handley}, 228

\bibitem[{{Borlaff} {et~al.}(2016){Borlaff}, {Eliche-Moral}, {Beckman}, \&
  {Font}}]{Borlaff2016}
{Borlaff}, A., {Eliche-Moral}, M.~C., {Beckman}, J., \& {Font}, J. 2016, \aap,
  591, L7, \dodoi{10.1051/0004-6361/201628868}

\bibitem[{{Borlaff} {et~al.}(2017){Borlaff}, {Eliche-Moral}, {Beckman},
  {Ciambur}, {P{\'e}rez-Gonz{\'a}lez}, {Barro}, {Cava}, \&
  {Cardiel}}]{Borlaff2017}
{Borlaff}, A., {Eliche-Moral}, M.~C., {Beckman}, J.~E., {et~al.} 2017, \aap,
  604, A119, \dodoi{10.1051/0004-6361/201630282}

\bibitem[{{Borlaff} {et~al.}(2019){Borlaff}, {Trujillo}, {Rom{\'a}n},
  {Beckman}, {Eliche-Moral}, {Infante-S{\'a}inz}, {Lumbreras-Calle}, {de
  Almagro}, {G{\'o}mez-Guijarro}, {Cebri{\'a}n}, {Dorta}, {Cardiel},
  {Akhlaghi}, \& {Mart{\'\i}nez-Lombilla}}]{Borlaff2019}
{Borlaff}, A., {Trujillo}, I., {Rom{\'a}n}, J., {et~al.} 2019, \aap, 621, A133,
  \dodoi{10.1051/0004-6361/201834312}

\bibitem[{{Bottema} {et~al.}(1987){Bottema}, {Shostak}, \& {van der
  Kruit}}]{Bottema1987}
{Bottema}, R., {Shostak}, G.~S., \& {van der Kruit}, P.~C. 1987, \nat, 328,
  401, \dodoi{10.1038/328401a0}

\bibitem[{{Brinks} \& {Burton}(1984)}]{Brinks1984}
{Brinks}, E., \& {Burton}, W.~B. 1984, \aap, 141, 195

\bibitem[{{de Grijs} \& {Peletier}(1997)}]{deGrijs1997}
{de Grijs}, R., \& {Peletier}, R.~F. 1997, \aap, 320, L21.
\newblock \doarXiv{astro-ph/9702215}

\bibitem[{{Eisenstein} {et~al.}(2011){Eisenstein}, {Weinberg}, {Agol},
  {Aihara}, {Allende Prieto}, {Anderson}, {Arns}, {Aubourg}, {Bailey},
  {Balbinot}, {Barkhouser}, {Beers}, {Berlind}, {Bickerton}, {Bizyaev},
  {Blanton}, {Bochanski}, {Bolton}, {Bosman}, {Bovy}, {Brandt}, {Breslauer},
  {Brewington}, {Brinkmann}, {Brown}, {Brownstein}, {Burger}, {Busca},
  {Campbell}, {Cargile}, {Carithers}, {Carlberg}, {Carr}, {Chang}, {Chen},
  {Chiappini}, {Comparat}, {Connolly}, {Cortes}, {Croft}, {Cunha}, {da Costa},
  {Davenport}, {Dawson}, {De Lee}, {Porto de Mello}, {de Simoni}, {Dean},
  {Dhital}, {Ealet}, {Ebelke}, {Edmondson}, {Eiting}, {Escoffier}, {Esposito},
  {Evans}, {Fan}, {Femen{\'\i}a Castell{\'a}}, {Dutra Ferreira}, {Fitzgerald},
  {Fleming}, {Font-Ribera}, {Ford}, {Frinchaboy}, {Garc{\'\i}a P{\'e}rez},
  {Gaudi}, {Ge}, {Ghezzi}, {Gillespie}, {Gilmore}, {Girardi}, {Gott}, {Gould},
  {Grebel}, {Gunn}, {Hamilton}, {Harding}, {Harris}, {Hawley}, {Hearty},
  {Hennawi}, {Gonz{\'a}lez Hern{\'a}ndez}, {Ho}, {Hogg}, {Holtzman},
  {Honscheid}, {Inada}, {Ivans}, {Jiang}, {Jiang}, {Johnson}, {Jordan},
  {Jordan}, {Kauffmann}, {Kazin}, {Kirkby}, {Klaene}, {Knapp}, {Kneib},
  {Kochanek}, {Koesterke}, {Kollmeier}, {Kron}, {Lampeitl}, {Lang}, {Lawler},
  {Le Goff}, {Lee}, {Lee}, {Leisenring}, {Lin}, {Liu}, {Long}, {Loomis},
  {Lucatello}, {Lundgren}, {Lupton}, {Ma}, {Ma}, {MacDonald}, {Mack},
  {Mahadevan}, {Maia}, {Majewski}, {Makler}, {Malanushenko}, {Malanushenko},
  {Mandelbaum}, {Maraston}, {Margala}, {Maseman}, {Masters}, {McBride},
  {McDonald}, {McGreer}, {McMahon}, {Mena Requejo}, {M{\'e}nard},
  {Miralda-Escud{\'e}}, {Morrison}, {Mullally}, {Muna}, {Murayama}, {Myers},
  {Naugle}, {Neto}, {Nguyen}, {Nichol}, {Nidever}, {O'Connell}, {Ogando},
  {Olmstead}, {Oravetz}, {Padmanabhan}, {Paegert}, {Palanque-Delabrouille},
  {Pan}, {Pandey}, {Parejko}, {P{\^a}ris}, {Pellegrini}, {Pepper}, {Percival},
  {Petitjean}, {Pfaffenberger}, {Pforr}, {Phleps}, {Pichon}, {Pieri}, {Prada},
  {Price-Whelan}, {Raddick}, {Ramos}, {Reid}, {Reyle}, {Rich}, {Richards},
  {Rieke}, {Rieke}, {Rix}, {Robin}, {Rocha-Pinto}, {Rockosi}, {Roe},
  {Rollinde}, {Ross}, {Ross}, {Rossetto}, {S{\'a}nchez}, {Santiago}, {Sayres},
  {Schiavon}, {Schlegel}, {Schlesinger}, {Schmidt}, {Schneider}, {Sellgren},
  {Shelden}, {Sheldon}, {Shetrone}, {Shu}, {Silverman}, {Simmerer}, {Simmons},
  {Sivarani}, {Skrutskie}, {Slosar}, {Smee}, {Smith}, {Snedden}, {Stassun},
  {Steele}, {Steinmetz}, {Stockett}, {Stollberg}, {Strauss}, {Szalay},
  {Tanaka}, {Thakar}, {Thomas}, {Tinker}, {Tofflemire}, {Tojeiro}, {Tremonti},
  {Vargas Maga{\~n}a}, {Verde}, {Vogt}, {Wake}, {Wan}, {Wang}, {Weaver},
  {White}, {White}, {Wilson}, {Wisniewski}, {Wood-Vasey}, {Yanny}, {Yasuda},
  {Y{\`e}che}, {York}, {Young}, {Zasowski}, {Zehavi}, \&
  {Zhao}}]{Eisenstein2011}
{Eisenstein}, D.~J., {Weinberg}, D.~H., {Agol}, E., {et~al.} 2011, \aj, 142,
  72, \dodoi{10.1088/0004-6256/142/3/72}

\bibitem[{{Erwin}(2015)}]{Erwin2015}
{Erwin}, P. 2015, \apj, 799, 226, \dodoi{10.1088/0004-637X/799/2/226}

\bibitem[{{Erwin} {et~al.}(2005){Erwin}, {Beckman}, \& {Pohlen}}]{Erwin2005}
{Erwin}, P., {Beckman}, J.~E., \& {Pohlen}, M. 2005, \apjl, 626, L81,
  \dodoi{10.1086/431739}

\bibitem[{{Erwin} {et~al.}(2008){Erwin}, {Pohlen}, \& {Beckman}}]{Erwin2008}
{Erwin}, P., {Pohlen}, M., \& {Beckman}, J.~E. 2008, \aj, 135, 20,
  \dodoi{10.1088/0004-6256/135/1/20}

\bibitem[{{Espada} {et~al.}(2011){Espada}, {Verdes-Montenegro}, {Huchtmeier},
  {Sulentic}, {Verley}, {Leon}, \& {Sabater}}]{2011yCat..35320117E}
{Espada}, D., {Verdes-Montenegro}, L., {Huchtmeier}, W.~K., {et~al.} 2011,
  VizieR Online Data Catalog, J/A+A/532/A117

\bibitem[{{Fliri} \& {Trujillo}(2016)}]{IACStripe82}
{Fliri}, J., \& {Trujillo}, I. 2016, \mnras, 456, 1359,
  \dodoi{10.1093/mnras/stv2686}

\bibitem[{{Freeman}(1970)}]{Freeman1970}
{Freeman}, K.~C. 1970, \apj, 160, 811, \dodoi{10.1086/150474}

\bibitem[{{Fritz} {et~al.}(2018){Fritz}, {Lokken}, {Kallivayalil}, {Wetzel},
  {Linden}, {Zivick}, \& {Tollerud}}]{Fritz2018ApJ860164F}
{Fritz}, T.~K., {Lokken}, M., {Kallivayalil}, N., {et~al.} 2018, \apj, 860,
  164, \dodoi{10.3847/1538-4357/aac516}

\bibitem[{{Gaia Collaboration} {et~al.}(2016){Gaia Collaboration}, {Prusti},
  {de Bruijne}, {Brown}, {Vallenari}, {Babusiaux}, {Bailer-Jones}, {Bastian},
  {Biermann}, {Evans}, {Eyer}, {Jansen}, {Jordi}, {Klioner}, {Lammers},
  {Lindegren}, {Luri}, {Mignard}, {Milligan}, {Panem}, {Poinsignon},
  {Pourbaix}, {Randich}, {Sarri}, {Sartoretti}, {Siddiqui}, {Soubiran},
  {Valette}, {van Leeuwen}, {Walton}, {Aerts}, {Arenou}, {Cropper}, {Drimmel},
  {H{\o}g}, {Katz}, {Lattanzi}, {O'Mullane}, {Grebel}, {Holland}, {Huc},
  {Passot}, {Bramante}, {Cacciari}, {Casta{\~n}eda}, {Chaoul}, {Cheek}, {De
  Angeli}, {Fabricius}, {Guerra}, {Hern{\'a}ndez}, {Jean-Antoine-Piccolo},
  {Masana}, {Messineo}, {Mowlavi}, {Nienartowicz}, {Ord{\'o}{\~n}ez-Blanco},
  {Panuzzo}, {Portell}, {Richards}, {Riello}, {Seabroke}, {Tanga},
  {Th{\'e}venin}, {Torra}, {Els}, {Gracia-Abril}, {Comoretto},
  {Garcia-Reinaldos}, {Lock}, {Mercier}, {Altmann}, {Andrae}, {Astraatmadja},
  {Bellas-Velidis}, {Benson}, {Berthier}, {Blomme}, {Busso}, {Carry},
  {Cellino}, {Clementini}, {Cowell}, {Creevey}, {Cuypers}, {Davidson}, {De
  Ridder}, {de Torres}, {Delchambre}, {Dell'Oro}, {Ducourant}, {Fr{\'e}mat},
  {Garc{\'\i}a-Torres}, {Gosset}, {Halbwachs}, {Hambly}, {Harrison}, {Hauser},
  {Hestroffer}, {Hodgkin}, {Huckle}, {Hutton}, {Jasniewicz}, {Jordan},
  {Kontizas}, {Korn}, {Lanzafame}, {Manteiga}, {Moitinho}, {Muinonen},
  {Osinde}, {Pancino}, {Pauwels}, {Petit}, {Recio-Blanco}, {Robin}, {Sarro},
  {Siopis}, {Smith}, {Smith}, {Sozzetti}, {Thuillot}, {van Reeven}, {Viala},
  {Abbas}, {Abreu Aramburu}, {Accart}, {Aguado}, {Allan}, {Allasia},
  {Altavilla}, {{\'A}lvarez}, {Alves}, {Anderson}, {Andrei}, {Anglada Varela},
  {Antiche}, {Antoja}, {Ant{\'o}n}, {Arcay}, {Atzei}, {Ayache}, {Bach},
  {Baker}, {Balaguer-N{\'u}{\~n}ez}, {Barache}, {Barata}, {Barbier}, {Barblan},
  {Baroni}, {Barrado y Navascu{\'e}s}, {Barros}, {Barstow}, {Becciani},
  {Bellazzini}, {Bellei}, {Bello Garc{\'\i}a}, {Belokurov}, {Bendjoya},
  {Berihuete}, {Bianchi}, {Bienaym{\'e}}, {Billebaud}, {Blagorodnova},
  {Blanco-Cuaresma}, {Boch}, {Bombrun}, {Borrachero}, {Bouquillon}, {Bourda},
  {Bouy}, {Bragaglia}, {Breddels}, {Brouillet}, {Br{\"u}semeister},
  {Bucciarelli}, {Budnik}, {Burgess}, {Burgon}, {Burlacu}, {Busonero}, {Buzzi},
  {Caffau}, {Cambras}, {Campbell}, {Cancelliere}, {Cantat-Gaudin}, {Carlucci},
  {Carrasco}, {Castellani}, {Charlot}, {Charnas}, {Charvet}, {Chassat},
  {Chiavassa}, {Clotet}, {Cocozza}, {Collins}, {Collins}, {Costigan}, {Crifo},
  {Cross}, {Crosta}, {Crowley}, {Dafonte}, {Damerdji}, {Dapergolas}, {David},
  {David}, {De Cat}, {de Felice}, {de Laverny}, {De Luise}, {De March}, {de
  Martino}, {de Souza}, {Debosscher}, {del Pozo}, {Delbo}, {Delgado},
  {Delgado}, {di Marco}, {Di Matteo}, {Diakite}, {Distefano}, {Dolding}, {Dos
  Anjos}, {Drazinos}, {Dur{\'a}n}, {Dzigan}, {Ecale}, {Edvardsson}, {Enke},
  {Erdmann}, {Escolar}, {Espina}, {Evans}, {Eynard Bontemps}, {Fabre},
  {Fabrizio}, {Faigler}, {Falc{\~a}o}, {Farr{\`a}s Casas}, {Faye}, {Federici},
  {Fedorets}, {Fern{\'a}ndez-Hern{\'a}ndez}, {Fernique}, {Fienga}, {Figueras},
  {Filippi}, {Findeisen}, {Fonti}, {Fouesneau}, {Fraile}, {Fraser}, {Fuchs},
  {Furnell}, {Gai}, {Galleti}, {Galluccio}, {Garabato}, {Garc{\'\i}a-Sedano},
  {Gar{\'e}}, {Garofalo}, {Garralda}, {Gavras}, {Gerssen}, {Geyer}, {Gilmore},
  {Girona}, {Giuffrida}, {Gomes}, {Gonz{\'a}lez-Marcos},
  {Gonz{\'a}lez-N{\'u}{\~n}ez}, {Gonz{\'a}lez-Vidal}, {Granvik}, {Guerrier},
  {Guillout}, {Guiraud}, {G{\'u}rpide}, {Guti{\'e}rrez-S{\'a}nchez}, {Guy},
  {Haigron}, {Hatzidimitriou}, {Haywood}, {Heiter}, {Helmi}, {Hobbs},
  {Hofmann}, {Holl}, {Holland}, {Hunt}, {Hypki}, {Icardi}, {Irwin}, {Jevardat
  de Fombelle}, {Jofr{\'e}}, {Jonker}, {Jorissen}, {Julbe}, {Karampelas},
  {Kochoska}, {Kohley}, {Kolenberg}, {Kontizas}, {Koposov}, {Kordopatis},
  {Koubsky}, {Kowalczyk}, {Krone-Martins}, {Kudryashova}, {Kull}, {Bachchan},
  {Lacoste-Seris}, {Lanza}, {Lavigne}, {Le Poncin-Lafitte}, {Lebreton},
  {Lebzelter}, {Leccia}, {Leclerc}, {Lecoeur-Taibi}, {Lemaitre}, {Lenhardt},
  {Leroux}, {Liao}, {Licata}, {Lindstr{\o}m}, {Lister}, {Livanou}, {Lobel},
  {L{\"o}ffler}, {L{\'o}pez}, {Lopez-Lozano}, {Lorenz}, {Loureiro},
  {MacDonald}, {Magalh{\~a}es Fernandes}, {Managau}, {Mann}, {Mantelet},
  {Marchal}, {Marchant}, {Marconi}, {Marie}, {Marinoni}, {Marrese},
  {Marschalk{\'o}}, {Marshall}, {Mart{\'\i}n-Fleitas}, {Martino}, {Mary},
  {Matijevi{\v{c}}}, {Mazeh}, {McMillan}, {Messina}, {Mestre}, {Michalik},
  {Millar}, {Miranda}, {Molina}, {Molinaro}, {Molinaro}, {Moln{\'a}r},
  {Moniez}, {Montegriffo}, {Monteiro}, {Mor}, {Mora}, {Morbidelli}, {Morel},
  {Morgenthaler}, {Morley}, {Morris}, {Mulone}, {Muraveva}, {Musella},
  {Narbonne}, {Nelemans}, {Nicastro}, {Noval}, {Ord{\'e}novic},
  {Ordieres-Mer{\'e}}, {Osborne}, {Pagani}, {Pagano}, {Pailler}, {Palacin},
  {Palaversa}, {Parsons}, {Paulsen}, {Pecoraro}, {Pedrosa}, {Pentik{\"a}inen},
  {Pereira}, {Pichon}, {Piersimoni}, {Pineau}, {Plachy}, {Plum}, {Poujoulet},
  {Pr{\v{s}}a}, {Pulone}, {Ragaini}, {Rago}, {Rambaux}, {Ramos-Lerate},
  {Ranalli}, {Rauw}, {Read}, {Regibo}, {Renk}, {Reyl{\'e}}, {Ribeiro},
  {Rimoldini}, {Ripepi}, {Riva}, {Rixon}, {Roelens}, {Romero-G{\'o}mez},
  {Rowell}, {Royer}, {Rudolph}, {Ruiz-Dern}, {Sadowski}, {Sagrist{\`a}
  Sell{\'e}s}, {Sahlmann}, {Salgado}, {Salguero}, {Sarasso}, {Savietto},
  {Schnorhk}, {Schultheis}, {Sciacca}, {Segol}, {Segovia}, {Segransan},
  {Serpell}, {Shih}, {Smareglia}, {Smart}, {Smith}, {Solano}, {Solitro},
  {Sordo}, {Soria Nieto}, {Souchay}, {Spagna}, {Spoto}, {Stampa}, {Steele},
  {Steidelm{\"u}ller}, {Stephenson}, {Stoev}, {Suess}, {S{\"u}veges}, {Surdej},
  {Szabados}, {Szegedi-Elek}, {Tapiador}, {Taris}, {Tauran}, {Taylor},
  {Teixeira}, {Terrett}, {Tingley}, {Trager}, {Turon}, {Ulla}, {Utrilla},
  {Valentini}, {van Elteren}, {Van Hemelryck}, {van Leeuwen}, {Varadi},
  {Vecchiato}, {Veljanoski}, {Via}, {Vicente}, {Vogt}, {Voss}, {Votruba},
  {Voutsinas}, {Walmsley}, {Weiler}, {Weingrill}, {Werner}, {Wevers},
  {Whitehead}, {Wyrzykowski}, {Yoldas}, {{\v{Z}}erjal}, {Zucker}, {Zurbach},
  {Zwitter}, {Alecu}, {Allen}, {Allende Prieto}, {Amorim},
  {Anglada-Escud{\'e}}, {Arsenijevic}, {Azaz}, {Balm}, {Beck}, {Bernstein},
  {Bigot}, {Bijaoui}, {Blasco}, {Bonfigli}, {Bono}, {Boudreault}, {Bressan},
  {Brown}, {Brunet}, {Bunclark}, {Buonanno}, {Butkevich}, {Carret}, {Carrion},
  {Chemin}, {Ch{\'e}reau}, {Corcione}, {Darmigny}, {de Boer}, {de Teodoro}, {de
  Zeeuw}, {Delle Luche}, {Domingues}, {Dubath}, {Fodor}, {Fr{\'e}zouls},
  {Fries}, {Fustes}, {Fyfe}, {Gallardo}, {Gallegos}, {Gardiol}, {Gebran},
  {Gomboc}, {G{\'o}mez}, {Grux}, {Gueguen}, {Heyrovsky}, {Hoar}, {Iannicola},
  {Isasi Parache}, {Janotto}, {Joliet}, {Jonckheere}, {Keil}, {Kim},
  {Klagyivik}, {Klar}, {Knude}, {Kochukhov}, {Kolka}, {Kos}, {Kutka}, {Lainey},
  {LeBouquin}, {Liu}, {Loreggia}, {Makarov}, {Marseille}, {Martayan},
  {Martinez-Rubi}, {Massart}, {Meynadier}, {Mignot}, {Munari}, {Nguyen},
  {Nordlander}, {Ocvirk}, {O'Flaherty}, {Olias Sanz}, {Ortiz}, {Osorio},
  {Oszkiewicz}, {Ouzounis}, {Palmer}, {Park}, {Pasquato}, {Peltzer}, {Peralta},
  {P{\'e}turaud}, {Pieniluoma}, {Pigozzi}, {Poels}, {Prat}, {Prod'homme},
  {Raison}, {Rebordao}, {Risquez}, {Rocca-Volmerange}, {Rosen}, {Ruiz-Fuertes},
  {Russo}, {Sembay}, {Serraller Vizcaino}, {Short}, {Siebert}, {Silva},
  {Sinachopoulos}, {Slezak}, {Soffel}, {Sosnowska}, {Strai{\v{z}}ys}, {ter
  Linden}, {Terrell}, {Theil}, {Tiede}, {Troisi}, {Tsalmantza}, {Tur},
  {Vaccari}, {Vachier}, {Valles}, {Van Hamme}, {Veltz}, {Virtanen}, {Wallut},
  {Wichmann}, {Wilkinson}, {Ziaeepour}, \& {Zschocke}}]{Gaia2016}
{Gaia Collaboration}, {Prusti}, T., {de Bruijne}, J.~H.~J., {et~al.} 2016,
  \aap, 595, A1, \dodoi{10.1051/0004-6361/201629272}

\bibitem[{{Gaia Collaboration} {et~al.}(2021){Gaia Collaboration}, {Brown},
  {Vallenari}, {Prusti}, {de Bruijne}, {Babusiaux}, {Biermann}, {Creevey},
  {Evans}, {Eyer}, {Hutton}, {Jansen}, {Jordi}, {Klioner}, {Lammers},
  {Lindegren}, {Luri}, {Mignard}, {Panem}, {Pourbaix}, {Randich}, {Sartoretti},
  {Soubiran}, {Walton}, {Arenou}, {Bailer-Jones}, {Bastian}, {Cropper},
  {Drimmel}, {Katz}, {Lattanzi}, {van Leeuwen}, {Bakker}, {Cacciari},
  {Casta{\~n}eda}, {De Angeli}, {Ducourant}, {Fabricius}, {Fouesneau},
  {Fr{\'e}mat}, {Guerra}, {Guerrier}, {Guiraud}, {Jean-Antoine Piccolo},
  {Masana}, {Messineo}, {Mowlavi}, {Nicolas}, {Nienartowicz}, {Pailler},
  {Panuzzo}, {Riclet}, {Roux}, {Seabroke}, {Sordo}, {Tanga}, {Th{\'e}venin},
  {Gracia-Abril}, {Portell}, {Teyssier}, {Altmann}, {Andrae}, {Bellas-Velidis},
  {Benson}, {Berthier}, {Blomme}, {Brugaletta}, {Burgess}, {Busso}, {Carry},
  {Cellino}, {Cheek}, {Clementini}, {Damerdji}, {Davidson}, {Delchambre},
  {Dell'Oro}, {Fern{\'a}ndez-Hern{\'a}ndez}, {Galluccio}, {Garc{\'\i}a-Lario},
  {Garcia-Reinaldos}, {Gonz{\'a}lez-N{\'u}{\~n}ez}, {Gosset}, {Haigron},
  {Halbwachs}, {Hambly}, {Harrison}, {Hatzidimitriou}, {Heiter},
  {Hern{\'a}ndez}, {Hestroffer}, {Hodgkin}, {Holl}, {Jan{\ss}en}, {Jevardat de
  Fombelle}, {Jordan}, {Krone-Martins}, {Lanzafame}, {L{\"o}ffler}, {Lorca},
  {Manteiga}, {Marchal}, {Marrese}, {Moitinho}, {Mora}, {Muinonen}, {Osborne},
  {Pancino}, {Pauwels}, {Petit}, {Recio-Blanco}, {Richards}, {Riello},
  {Rimoldini}, {Robin}, {Roegiers}, {Rybizki}, {Sarro}, {Siopis}, {Smith},
  {Sozzetti}, {Ulla}, {Utrilla}, {van Leeuwen}, {van Reeven}, {Abbas}, {Abreu
  Aramburu}, {Accart}, {Aerts}, {Aguado}, {Ajaj}, {Altavilla}, {{\'A}lvarez},
  {{\'A}lvarez Cid-Fuentes}, {Alves}, {Anderson}, {Anglada Varela}, {Antoja},
  {Audard}, {Baines}, {Baker}, {Balaguer-N{\'u}{\~n}ez}, {Balbinot}, {Balog},
  {Barache}, {Barbato}, {Barros}, {Barstow}, {Bartolom{\'e}}, {Bassilana},
  {Bauchet}, {Baudesson-Stella}, {Becciani}, {Bellazzini}, {Bernet}, {Bertone},
  {Bianchi}, {Blanco-Cuaresma}, {Boch}, {Bombrun}, {Bossini}, {Bouquillon},
  {Bragaglia}, {Bramante}, {Breedt}, {Bressan}, {Brouillet}, {Bucciarelli},
  {Burlacu}, {Busonero}, {Butkevich}, {Buzzi}, {Caffau}, {Cancelliere},
  {C{\'a}novas}, {Cantat-Gaudin}, {Carballo}, {Carlucci}, {Carnerero},
  {Carrasco}, {Casamiquela}, {Castellani}, {Castro-Ginard}, {Castro Sampol},
  {Chaoul}, {Charlot}, {Chemin}, {Chiavassa}, {Cioni}, {Comoretto}, {Cooper},
  {Cornez}, {Cowell}, {Crifo}, {Crosta}, {Crowley}, {Dafonte}, {Dapergolas},
  {David}, {David}, {de Laverny}, {De Luise}, {De March}, {De Ridder}, {de
  Souza}, {de Teodoro}, {de Torres}, {del Peloso}, {del Pozo}, {Delbo},
  {Delgado}, {Delgado}, {Delisle}, {Di Matteo}, {Diakite}, {Diener},
  {Distefano}, {Dolding}, {Eappachen}, {Edvardsson}, {Enke}, {Esquej}, {Fabre},
  {Fabrizio}, {Faigler}, {Fedorets}, {Fernique}, {Fienga}, {Figueras},
  {Fouron}, {Fragkoudi}, {Fraile}, {Franke}, {Gai}, {Garabato},
  {Garcia-Gutierrez}, {Garc{\'\i}a-Torres}, {Garofalo}, {Gavras}, {Gerlach},
  {Geyer}, {Giacobbe}, {Gilmore}, {Girona}, {Giuffrida}, {Gomel}, {Gomez},
  {Gonzalez-Santamaria}, {Gonz{\'a}lez-Vidal}, {Granvik},
  {Guti{\'e}rrez-S{\'a}nchez}, {Guy}, {Hauser}, {Haywood}, {Helmi}, {Hidalgo},
  {Hilger}, {H{\l}adczuk}, {Hobbs}, {Holland}, {Huckle}, {Jasniewicz},
  {Jonker}, {Juaristi Campillo}, {Julbe}, {Karbevska}, {Kervella}, {Khanna},
  {Kochoska}, {Kontizas}, {Kordopatis}, {Korn}, {Kostrzewa-Rutkowska},
  {Kruszy{\'n}ska}, {Lambert}, {Lanza}, {Lasne}, {Le Campion}, {Le Fustec},
  {Lebreton}, {Lebzelter}, {Leccia}, {Leclerc}, {Lecoeur-Taibi}, {Liao},
  {Licata}, {Lindstr{\o}m}, {Lister}, {Livanou}, {Lobel}, {Madrero Pardo},
  {Managau}, {Mann}, {Marchant}, {Marconi}, {Marcos Santos}, {Marinoni},
  {Marocco}, {Marshall}, {Martin Polo}, {Mart{\'\i}n-Fleitas}, {Masip},
  {Massari}, {Mastrobuono-Battisti}, {Mazeh}, {McMillan}, {Messina},
  {Michalik}, {Millar}, {Mints}, {Molina}, {Molinaro}, {Moln{\'a}r},
  {Montegriffo}, {Mor}, {Morbidelli}, {Morel}, {Morris}, {Mulone}, {Munoz},
  {Muraveva}, {Murphy}, {Musella}, {Noval}, {Ord{\'e}novic}, {Orr{\`u}},
  {Osinde}, {Pagani}, {Pagano}, {Palaversa}, {Palicio}, {Panahi}, {Pawlak},
  {Pe{\~n}alosa Esteller}, {Penttil{\"a}}, {Piersimoni}, {Pineau}, {Plachy},
  {Plum}, {Poggio}, {Poretti}, {Poujoulet}, {Pr{\v{s}}a}, {Pulone}, {Racero},
  {Ragaini}, {Rainer}, {Raiteri}, {Rambaux}, {Ramos}, {Ramos-Lerate}, {Re
  Fiorentin}, {Regibo}, {Reyl{\'e}}, {Ripepi}, {Riva}, {Rixon}, {Robichon},
  {Robin}, {Roelens}, {Rohrbasser}, {Romero-G{\'o}mez}, {Rowell}, {Royer},
  {Rybicki}, {Sadowski}, {Sagrist{\`a} Sell{\'e}s}, {Sahlmann}, {Salgado},
  {Salguero}, {Samaras}, {Sanchez Gimenez}, {Sanna}, {Santove{\~n}a},
  {Sarasso}, {Schultheis}, {Sciacca}, {Segol}, {Segovia}, {S{\'e}gransan},
  {Semeux}, {Shahaf}, {Siddiqui}, {Siebert}, {Siltala}, {Slezak}, {Smart},
  {Solano}, {Solitro}, {Souami}, {Souchay}, {Spagna}, {Spoto}, {Steele},
  {Steidelm{\"u}ller}, {Stephenson}, {S{\"u}veges}, {Szabados}, {Szegedi-Elek},
  {Taris}, {Tauran}, {Taylor}, {Teixeira}, {Thuillot}, {Tonello}, {Torra},
  {Torra}, {Turon}, {Unger}, {Vaillant}, {van Dillen}, {Vanel}, {Vecchiato},
  {Viala}, {Vicente}, {Voutsinas}, {Weiler}, {Wevers}, {Wyrzykowski}, {Yoldas},
  {Yvard}, {Zhao}, {Zorec}, {Zucker}, {Zurbach}, \& {Zwitter}}]{Gaia2021}
{Gaia Collaboration}, {Brown}, A.~G.~A., {Vallenari}, A., {et~al.} 2021, \aap,
  649, A1, \dodoi{10.1051/0004-6361/202039657}

\bibitem[{{Giavalisco} {et~al.}(2004){Giavalisco}, {Ferguson}, {Koekemoer},
  {Dickinson}, {Alexander}, {Bauer}, {Bergeron}, {Biagetti}, {Brandt},
  {Casertano}, {Cesarsky}, {Chatzichristou}, {Conselice}, {Cristiani}, {Da
  Costa}, {Dahlen}, {de Mello}, {Eisenhardt}, {Erben}, {Fall}, {Fassnacht},
  {Fosbury}, {Fruchter}, {Gardner}, {Grogin}, {Hook}, {Hornschemeier}, {Idzi},
  {Jogee}, {Kretchmer}, {Laidler}, {Lee}, {Livio}, {Lucas}, {Madau},
  {Mobasher}, {Moustakas}, {Nonino}, {Padovani}, {Papovich}, {Park},
  {Ravindranath}, {Renzini}, {Richardson}, {Riess}, {Rosati}, {Schirmer},
  {Schreier}, {Somerville}, {Spinrad}, {Stern}, {Stiavelli}, {Strolger},
  {Urry}, {Vandame}, {Williams}, \& {Wolf}}]{Giavalisco2004}
{Giavalisco}, M., {Ferguson}, H.~C., {Koekemoer}, A.~M., {et~al.} 2004, \apjl,
  600, L93, \dodoi{10.1086/379232}

\bibitem[{Harris {et~al.}(2020)Harris, Millman, van~der Walt, Gommers,
  Virtanen, Cournapeau, Wieser, Taylor, Berg, Smith, Kern, Picus, Hoyer, van
  Kerkwijk, Brett, Haldane, del R{\'{i}}o, Wiebe, Peterson,
  G{\'{e}}rard-Marchant, Sheppard, Reddy, Weckesser, Abbasi, Gohlke, \&
  Oliphant}]{harris2020array}
Harris, C.~R., Millman, K.~J., van~der Walt, S.~J., {et~al.} 2020, Nature, 585,
  357, \dodoi{10.1038/s41586-020-2649-2}

\bibitem[{{Haynes} {et~al.}(2018{\natexlab{a}}){Haynes}, {Giovanelli}, {Kent},
  {Adams}, {Balonek}, {Craig}, {Fertig}, {Finn}, {Giovanardi}, {Hallenbeck},
  {Hess}, {Hoffman}, {Huang}, {Jones}, {Koopmann}, {Kornreich}, {Leisman},
  {Miller}, {Moorman}, {O'Connor}, {O'Donoghue}, {Papastergis}, {Troischt},
  {Stark}, \& {Xiao}}]{Haynes2018}
{Haynes}, M.~P., {Giovanelli}, R., {Kent}, B.~R., {et~al.} 2018{\natexlab{a}},
  \apj, 861, 49, \dodoi{10.3847/1538-4357/aac956}

\bibitem[{{Haynes} {et~al.}(2018{\natexlab{b}}){Haynes}, {Giovanelli}, {Kent},
  {Adams}, {Balonek}, {Craig}, {Fertig}, {Finn}, {Giovanardi}, {Hallenbeck},
  {Hess}, {Hoffman}, {Huang}, {Jones}, {Koopmann}, {Kornreich}, {Leisman},
  {Miller}, {Moorman}, {O'Connor}, {O'Donoghue}, {Papastergis}, {Troischt},
  {Stark}, \& {Xiao}}]{Haynes2018ApJ86149H}
---. 2018{\natexlab{b}}, \apj, 861, 49, \dodoi{10.3847/1538-4357/aac956}

\bibitem[{{Hewitt} {et~al.}(1983){Hewitt}, {Haynes}, \&
  {Giovanelli}}]{1983AJ.....88..272H}
{Hewitt}, J.~N., {Haynes}, M.~P., \& {Giovanelli}, R. 1983, \aj, 88, 272,
  \dodoi{10.1086/113317}

\bibitem[{Hunter(2007)}]{Hunter:2007}
Hunter, J.~D. 2007, Computing in Science \& Engineering, 9, 90,
  \dodoi{10.1109/MCSE.2007.55}

\bibitem[{{Ibata} {et~al.}(1995){Ibata}, {Gilmore}, \&
  {Irwin}}]{Ibata1995MNRAS277781I}
{Ibata}, R.~A., {Gilmore}, G., \& {Irwin}, M.~J. 1995, \mnras, 277, 781,
  \dodoi{10.1093/mnras/277.3.781}

\bibitem[{{Infante-Sainz} {et~al.}(2020){Infante-Sainz}, {Trujillo}, \&
  {Rom{\'a}n}}]{InfanteSainz2020}
{Infante-Sainz}, R., {Trujillo}, I., \& {Rom{\'a}n}, J. 2020, \mnras, 491,
  5317, \dodoi{10.1093/mnras/stz3111}

\bibitem[{{James} {et~al.}(2008){James}, {Knapen}, {Shane}, {Baldry}, \& {de
  Jong}}]{James2008}
{James}, P.~A., {Knapen}, J.~H., {Shane}, N.~S., {Baldry}, I.~K., \& {de Jong},
  R.~S. 2008, \aap, 482, 507, \dodoi{10.1051/0004-6361:20078560}

\bibitem[{{Kennicutt}(1989)}]{Kennicutt1989}
{Kennicutt}, Robert~C., J. 1989, \apj, 344, 685, \dodoi{10.1086/167834}

\bibitem[{{Koda} {et~al.}(2015){Koda}, {Yagi}, {Yamanoi}, \&
  {Komiyama}}]{Koda2015ApJ807L2K}
{Koda}, J., {Yagi}, M., {Yamanoi}, H., \& {Komiyama}, Y. 2015, \apjl, 807, L2,
  \dodoi{10.1088/2041-8205/807/1/L2}

\bibitem[{{Kregel} {et~al.}(2002){Kregel}, {van der Kruit}, \& {de
  Grijs}}]{Kregel2002MNRAS.334..646K}
{Kregel}, M., {van der Kruit}, P.~C., \& {de Grijs}, R. 2002, \mnras, 334, 646,
  \dodoi{10.1046/j.1365-8711.2002.05556.x}

\bibitem[{{Kuijken} \& {Garcia-Ruiz}(2001)}]{Kuijken2001}
{Kuijken}, K., \& {Garcia-Ruiz}, I. 2001, \aspconf, Vol. 230, {Galactic Disk
  Warps}, ed. J.~G. {Funes} \& E.~M. {Corsini}, 401--408

\bibitem[{{Laporte} {et~al.}(2019){Laporte}, {Minchev}, {Johnston}, \&
  {G{\'o}mez}}]{Laporte2019MNRAS.485.3134L}
{Laporte}, C. F.~P., {Minchev}, I., {Johnston}, K.~V., \& {G{\'o}mez}, F.~A.
  2019, \mnras, 485, 3134, \dodoi{10.1093/mnras/stz583}

\bibitem[{{Martin} {et~al.}(2013){Martin}, {Ibata}, {McConnachie}, {Mackey},
  {Ferguson}, {Irwin}, {Lewis}, \& {Fardal}}]{Martin2013ApJ77680M}
{Martin}, N.~F., {Ibata}, R.~A., {McConnachie}, A.~W., {et~al.} 2013, \apj,
  776, 80, \dodoi{10.1088/0004-637X/776/2/80}

\bibitem[{{Mart{\'\i}n-Navarro} {et~al.}(2012){Mart{\'\i}n-Navarro}, {Bakos},
  {Trujillo}, {Knapen}, {Athanassoula}, {Bosma}, {Comer{\'o}n}, {Elmegreen},
  {Erroz-Ferrer}, {Gadotti}, {Gil de Paz}, {Hinz}, {Ho}, {Holwerda}, {Kim},
  {Laine}, {Laurikainen}, {Men{\'e}ndez-Delmestre}, {Mizusawa},
  {Mu{\~n}oz-Mateos}, {Regan}, {Salo}, {Seibert}, \&
  {Sheth}}]{MartinNavarro2012}
{Mart{\'\i}n-Navarro}, I., {Bakos}, J., {Trujillo}, I., {et~al.} 2012, \mnras,
  427, 1102, \dodoi{10.1111/j.1365-2966.2012.21929.x}

\bibitem[{{Mart{\'\i}nez-Delgado} {et~al.}(2009){Mart{\'\i}nez-Delgado},
  {Pohlen}, {Gabany}, {Majewski}, {Pe{\~n}arrubia}, \&
  {Palma}}]{MartinezDelgado2009ApJ692955M}
{Mart{\'\i}nez-Delgado}, D., {Pohlen}, M., {Gabany}, R.~J., {et~al.} 2009,
  \apj, 692, 955, \dodoi{10.1088/0004-637X/692/2/955}

\bibitem[{{Mart{\'\i}nez-Lombilla} \&
  {Knapen}(2019)}]{martinezlombilla+2019aap629_12}
{Mart{\'\i}nez-Lombilla}, C., \& {Knapen}, J.~H. 2019, \aap, 629, A12,
  \dodoi{10.1051/0004-6361/201935464}

\bibitem[{{Mathis}(1990)}]{Mathis1990ARA&A..28...37M}
{Mathis}, J.~S. 1990, \araa, 28, 37,
  \dodoi{10.1146/annurev.aa.28.090190.000345}

\bibitem[{Mestel(1963)}]{10.1093/mnras/126.6.553}
Mestel, L. 1963, Monthly Notices of the Royal Astronomical Society, 126, 553,
  \dodoi{10.1093/mnras/126.6.553}

\bibitem[{{O'Brien} {et~al.}(2010){O'Brien}, {Freeman}, {van der Kruit}, \&
  {Bosma}}]{OBrien2010}
{O'Brien}, J.~C., {Freeman}, K.~C., {van der Kruit}, P.~C., \& {Bosma}, A.
  2010, \aap, 515, A60, \dodoi{10.1051/0004-6361/200912565}

\bibitem[{Oke(1971)}]{Oke1971}
Oke, J.~B. 1971, ApJ, 170, 193, \dodoi{10.1086/151202}

\bibitem[{{Parnovsky} \& {Parnowski}(2010)}]{2010Ap&SS.325..163P}
{Parnovsky}, S.~L., \& {Parnowski}, A.~S. 2010, \apss, 325, 163,
  \dodoi{10.1007/s10509-009-0176-6}

\bibitem[{{Poggio} {et~al.}(2020){Poggio}, {Drimmel}, {Andrae}, {Bailer-Jones},
  {Fouesneau}, {Lattanzi}, {Smart}, \& {Spagna}}]{Poggio2020NatAs...4..590P}
{Poggio}, E., {Drimmel}, R., {Andrae}, R., {et~al.} 2020, Nature Astronomy, 4,
  590, \dodoi{10.1038/s41550-020-1017-3}

\bibitem[{{Pohlen} {et~al.}(2004){Pohlen}, {Balcells}, {L{\"u}tticke}, \&
  {Dettmar}}]{Pohlen2004}
{Pohlen}, M., {Balcells}, M., {L{\"u}tticke}, R., \& {Dettmar}, R.~J. 2004,
  \aap, 422, 465, \dodoi{10.1051/0004-6361:20035932}

\bibitem[{{Pohlen} {et~al.}(2002){Pohlen}, {Dettmar}, {L{\"u}tticke}, \&
  {Aronica}}]{Pohlen2002}
{Pohlen}, M., {Dettmar}, R.~J., {L{\"u}tticke}, R., \& {Aronica}, G. 2002,
  \aap, 392, 807, \dodoi{10.1051/0004-6361:20020994}

\bibitem[{{Pohlen} \& {Trujillo}(2006)}]{Pohlen2006}
{Pohlen}, M., \& {Trujillo}, I. 2006, \aap, 454, 759,
  \dodoi{10.1051/0004-6361:20064883}

\bibitem[{{Pohlen} {et~al.}(2007){Pohlen}, {Zaroubi}, {Peletier}, \&
  {Dettmar}}]{Pohlen2007}
{Pohlen}, M., {Zaroubi}, S., {Peletier}, R.~F., \& {Dettmar}, R.-J. 2007,
  \mnras, 378, 594, \dodoi{10.1111/j.1365-2966.2007.11790.x}

\bibitem[{{Rogstad} {et~al.}(1974){Rogstad}, {Lockhart}, \&
  {Wright}}]{Rogstad1974ApJ...193..309R}
{Rogstad}, D.~H., {Lockhart}, I.~A., \& {Wright}, M.~C.~H. 1974, \apj, 193,
  309, \dodoi{10.1086/153164}

\bibitem[{{Rom{\'a}n} {et~al.}(2021){Rom{\'a}n}, {Castilla}, \&
  {Pascual-Granado}}]{Roman2021A&A656A44R}
{Rom{\'a}n}, J., {Castilla}, A., \& {Pascual-Granado}, J. 2021, \aap, 656, A44,
  \dodoi{10.1051/0004-6361/202142161}

\bibitem[{{Rom{\'a}n} \& {Trujillo}(2018)}]{Roman2018}
{Rom{\'a}n}, J., \& {Trujillo}, I. 2018, Research Notes of the American
  Astronomical Society, 2, 144, \dodoi{10.3847/2515-5172/aad8b8}

\bibitem[{{Ro{\v{s}}kar} {et~al.}(2008){Ro{\v{s}}kar}, {Debattista}, {Stinson},
  {Quinn}, {Kaufmann}, \& {Wadsley}}]{Roskar2008}
{Ro{\v{s}}kar}, R., {Debattista}, V.~P., {Stinson}, G.~S., {et~al.} 2008,
  \apjl, 675, L65, \dodoi{10.1086/586734}

\bibitem[{{Sales} {et~al.}(2020){Sales}, {Navarro}, {Pe{\~n}afiel}, {Peng},
  {Lim}, \& {Hernquist}}]{Sales2020MNRAS4941848S}
{Sales}, L.~V., {Navarro}, J.~F., {Pe{\~n}afiel}, L., {et~al.} 2020, \mnras,
  494, 1848, \dodoi{10.1093/mnras/staa854}

\bibitem[{{Sancisi} \& {Allen}(1979)}]{Sancisi1979}
{Sancisi}, R., \& {Allen}, R.~J. 1979, \aap, 74, 73

\bibitem[{{Sandin}(2014)}]{Sandin2014}
{Sandin}, C. 2014, \aap, 567, A97, \dodoi{10.1051/0004-6361/201423429}

\bibitem[{{Sandin}(2015)}]{Sandin2015}
---. 2015, \aap, 577, A106, \dodoi{10.1051/0004-6361/201425168}

\bibitem[{{Shen} \& {Sellwood}(2006)}]{Shen2006MNRAS.370....2S}
{Shen}, J., \& {Sellwood}, J.~A. 2006, \mnras, 370, 2,
  \dodoi{10.1111/j.1365-2966.2006.10477.x}

\bibitem[{{Skowron} {et~al.}(2019){Skowron}, {Skowron}, {Mr{\'o}z}, {Udalski},
  {Pietrukowicz}, {Soszy{\'n}ski}, {Szyma{\'n}ski}, {Poleski}, {Koz{\l}owski},
  {Ulaczyk}, {Rybicki}, \& {Iwanek}}]{Skowron2019Sci...365..478S}
{Skowron}, D.~M., {Skowron}, J., {Mr{\'o}z}, P., {et~al.} 2019, Science, 365,
  478, \dodoi{10.1126/science.aau3181}

\bibitem[{Sparke \& Casertano(1988)}]{Sparke1988.10.1093mnras234.4.873}
Sparke, L.~S., \& Casertano, S. 1988, Monthly Notices of the Royal Astronomical
  Society, 234, 873, \dodoi{10.1093/mnras/234.4.873}

\bibitem[{{Spergel} {et~al.}(2007){Spergel}, {Bean}, {Dor{\'e}}, {Nolta},
  {Bennett}, {Dunkley}, {Hinshaw}, {Jarosik}, {Komatsu}, {Page}, {Peiris},
  {Verde}, {Halpern}, {Hill}, {Kogut}, {Limon}, {Meyer}, {Odegard}, {Tucker},
  {Weiland}, {Wollack}, \& {Wright}}]{WMAP2007}
{Spergel}, D.~N., {Bean}, R., {Dor{\'e}}, O., {et~al.} 2007, \apjs, 170, 377,
  \dodoi{10.1086/513700}

\bibitem[{{The Astropy Collaboration} {et~al.}(2013){The Astropy
  Collaboration}, {Robitaille, Thomas P.}, {Tollerud, Erik J.}, {Greenfield,
  Perry}, {Droettboom, Michael}, {Bray, Erik}, {Aldcroft, Tom}, {Davis, Matt},
  {Ginsburg, Adam}, {Price-Whelan, Adrian M.}, {Kerzendorf, Wolfgang E.},
  {Conley, Alexander}, {Crighton, Neil}, {Barbary, Kyle}, {Muna, Demitri},
  {Ferguson, Henry}, {Grollier, Fr\'ed\'eric}, {Parikh, Madhura M.}, {Nair,
  Prasanth H.}, {G\"unther, Hans M.}, {Deil, Christoph}, {Woillez, Julien},
  {Conseil, Simon}, {Kramer, Roban}, {Turner, James E. H.}, {Singer, Leo},
  {Fox, Ryan}, {Weaver, Benjamin A.}, {Zabalza, Victor}, {Edwards, Zachary I.},
  {Azalee Bostroem, K.}, {Burke, D. J.}, {Casey, Andrew R.}, {Crawford, Steven
  M.}, {Dencheva, Nadia}, {Ely, Justin}, {Jenness, Tim}, {Labrie, Kathleen},
  {Lim, Pey Lian}, {Pierfederici, Francesco}, {Pontzen, Andrew}, {Ptak, Andy},
  {Refsdal, Brian}, {Servillat, Mathieu}, \& {Streicher, Ole}}]{refId0}
{The Astropy Collaboration}, {Robitaille, Thomas P.}, {Tollerud, Erik J.},
  {et~al.} 2013, A\&A, 558, A33, \dodoi{10.1051/0004-6361/201322068}

\bibitem[{{Toomre}(1964)}]{Toomre1964}
{Toomre}, A. 1964, \apj, 139, 1217, \dodoi{10.1086/147861}

\bibitem[{{Trujillo} \& {Bakos}(2013)}]{Trujillo2013}
{Trujillo}, I., \& {Bakos}, J. 2013, \mnras, 431, 1121,
  \dodoi{10.1093/mnras/stt232}

\bibitem[{{Trujillo} \& {Fliri}(2016)}]{Trujillo2016}
{Trujillo}, I., \& {Fliri}, J. 2016, \apj, 823, 123,
  \dodoi{10.3847/0004-637X/823/2/123}

\bibitem[{{van der Kruit}(1979)}]{vanderkruit1979}
{van der Kruit}, P.~C. 1979, \aaps, 38, 15

\bibitem[{{van der Kruit}(2007)}]{vanderKruit2007A&A466883V}
---. 2007, \aap, 466, 883, \dodoi{10.1051/0004-6361:20066941}

\bibitem[{{van der Kruit} \& {Searle}(1981{\natexlab{a}})}]{vanderKruit1981a}
{van der Kruit}, P.~C., \& {Searle}, L. 1981{\natexlab{a}}, \aap, 95, 105

\bibitem[{{van der Kruit} \& {Searle}(1981{\natexlab{b}})}]{vanderKruit1981b}
---. 1981{\natexlab{b}}, \aap, 95, 116

\bibitem[{{Venhola} {et~al.}(2017){Venhola}, {Peletier}, {Laurikainen}, {Salo},
  {Lisker}, {Iodice}, {Capaccioli}, {Verdois Kleijn}, {Valentijn}, {Mieske},
  {Hilker}, {Wittmann}, {van de Ven}, {Grado}, {Spavone}, {Cantiello},
  {Napolitano}, {Paolillo}, \& {Falc{\'o}n-Barroso}}]{Venhola2017A&A608A142V}
{Venhola}, A., {Peletier}, R., {Laurikainen}, E., {et~al.} 2017, \aap, 608,
  A142, \dodoi{10.1051/0004-6361/201730696}

\bibitem[{{Weinberg} \& {Blitz}(2006)}]{Weinberg2006ApJ...641L..33W}
{Weinberg}, M.~D., \& {Blitz}, L. 2006, \apjl, 641, L33, \dodoi{10.1086/503607}

\bibitem[{{York} {et~al.}(2000){York}, {Adelman}, {Anderson}, {Anderson},
  {Annis}, {Bahcall}, {Bakken}, {Barkhouser}, {Bastian}, {Berman}, {Boroski},
  {Bracker}, {Briegel}, {Briggs}, {Brinkmann}, {Brunner}, {Burles}, {Carey},
  {Carr}, {Castander}, {Chen}, {Colestock}, {Connolly}, {Crocker}, {Csabai},
  {Czarapata}, {Davis}, {Doi}, {Dombeck}, {Eisenstein}, {Ellman}, {Elms},
  {Evans}, {Fan}, {Federwitz}, {Fiscelli}, {Friedman}, {Frieman}, {Fukugita},
  {Gillespie}, {Gunn}, {Gurbani}, {de Haas}, {Haldeman}, {Harris}, {Hayes},
  {Heckman}, {Hennessy}, {Hindsley}, {Holm}, {Holmgren}, {Huang}, {Hull},
  {Husby}, {Ichikawa}, {Ichikawa}, {Ivezi{\'c}}, {Kent}, {Kim}, {Kinney},
  {Klaene}, {Kleinman}, {Kleinman}, {Knapp}, {Korienek}, {Kron}, {Kunszt},
  {Lamb}, {Lee}, {Leger}, {Limmongkol}, {Lindenmeyer}, {Long}, {Loomis},
  {Loveday}, {Lucinio}, {Lupton}, {MacKinnon}, {Mannery}, {Mantsch}, {Margon},
  {McGehee}, {McKay}, {Meiksin}, {Merelli}, {Monet}, {Munn}, {Narayanan},
  {Nash}, {Neilsen}, {Neswold}, {Newberg}, {Nichol}, {Nicinski}, {Nonino},
  {Okada}, {Okamura}, {Ostriker}, {Owen}, {Pauls}, {Peoples}, {Peterson},
  {Petravick}, {Pier}, {Pope}, {Pordes}, {Prosapio}, {Rechenmacher}, {Quinn},
  {Richards}, {Richmond}, {Rivetta}, {Rockosi}, {Ruthmansdorfer}, {Sand ford},
  {Schlegel}, {Schneider}, {Sekiguchi}, {Sergey}, {Shimasaku}, {Siegmund},
  {Smee}, {Smith}, {Snedden}, {Stone}, {Stoughton}, {Strauss}, {Stubbs},
  {SubbaRao}, {Szalay}, {Szapudi}, {Szokoly}, {Thakar}, {Tremonti}, {Tucker},
  {Uomoto}, {Vanden Berk}, {Vogeley}, {Waddell}, {Wang}, {Watanabe},
  {Weinberg}, {Yanny}, {Yasuda}, \& {SDSS Collaboration}}]{York2000}
{York}, D.~G., {Adelman}, J., {Anderson}, John~E., J., {et~al.} 2000, \aj, 120,
  1579, \dodoi{10.1086/301513}

\bibitem[{{Zschaechner} {et~al.}(2012){Zschaechner}, {Rand}, {Heald},
  {Gentile}, \& {J{\'o}zsa}}]{Zschaechner2012}
{Zschaechner}, L.~K., {Rand}, R.~J., {Heald}, G.~H., {Gentile}, G., \&
  {J{\'o}zsa}, G. 2012, \apj, 760, 37, \dodoi{10.1088/0004-637X/760/1/37}

\end{thebibliography}

\clearpage
\appendix
\section{Appendix A}
\label{A1:flare_sides}
\begin{figure*}[h!]
 \begin{center}
 
\includegraphics[trim={0 0 0 0}, clip, width=\textwidth]{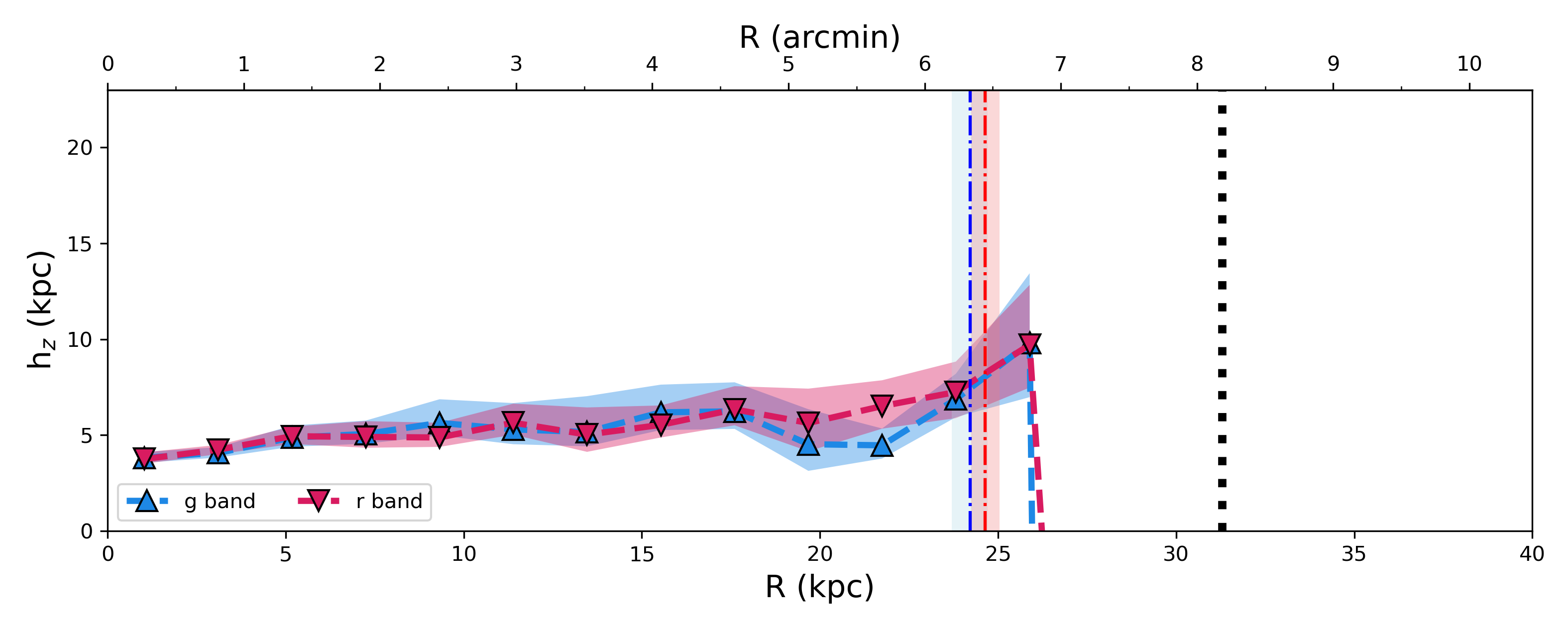}
\includegraphics[trim={0 0 0 0}, clip, width=\textwidth]{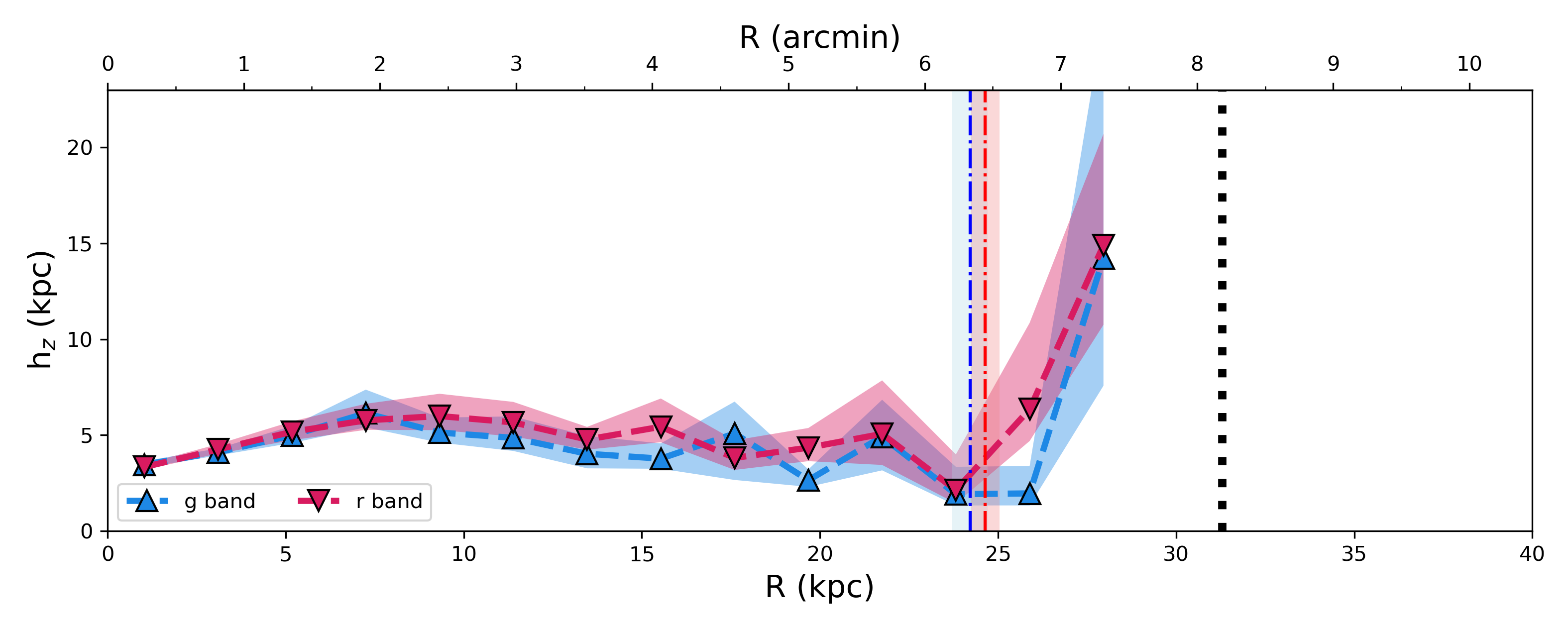}
\caption{Vertical disk scale-height (\hz) vs. galactocentric radius (R) profile. \emph{Top:} East side of \ugc. \emph{Bottom} West side of \ugc. Data are shown for the bands $g$ (\emph{blue} triangles) and $r$ (\emph{red} inverted triangles) associated with errors (\emph{blue} and \emph{red} shaded areas, respectively).}
\label{fig:flare_sides_galaxy}
\end{center}
\end{figure*}

In Fig.\,\ref{fig:flare_sides_galaxy} we show the analysis of the variation of the vertical scale-length as a function of the galactocentric radius, for the East and West side of \ugc\ galaxy.  The increase in \hz\ around $R=24$ kpc indicates the presence of a \emph{flare} in the disk of \ugc\ in both sides of the galaxy. Interestingly, the flare in the warped (East) side of the galaxy is less significative, indicating that the flare it is not an artifact in the surface brightness profiles created by the warp of the disk.

\end{document}